\pdfoutput=1

\documentclass{article}
\usepackage{amsmath,amssymb,amsfonts}
\usepackage{graphicx}
\usepackage{ifthen} 
\usepackage{xcolor}

\usepackage{geometry}
\usepackage{hyperref}

\newboolean{doubleBlind}
\setboolean{doubleBlind}{false}

\newboolean{removeDeletedText}
\setboolean{removeDeletedText}{true}

\definecolor{colourText}{rgb}{0,0,0}
\definecolor{colourRevisionContent}{rgb}{0,0,0}
\definecolor{colourRevisionText}{rgb}{0,0,0}
\definecolor{colourRevisionTodo}{rgb}{0,0,0}
\definecolor{colourRevisionDelete}{rgb}{0.8,0.8,0.8}

\ifthenelse{\boolean{removeDeletedText}}{
  \newcommand{\revisionDelete}[1]{}
}{
  \newcommand{\revisionDelete}[1]{\color{colourRevisionDelete} #1 \color{colourText}}
}

\author{
  D.E.~Charrier
  \and
  B.~Hazelwood
  \and 
  T.~Weinzierl\thanks{
    Department of Computer Science, Durham University, 
    Lower Mountjoy South Road, Durham DH1 3LE, United Kingdom 
    (\{tobias.weinzierl@durham.ac.uk\})
  }
}

\usepackage{ifthen} 
\usepackage{amsmath,amssymb,amsfonts}
\usepackage{graphicx}
\usepackage{algorithm}
\usepackage{algpseudocode}
\usepackage{color}
\usepackage{verbatim}
\usepackage[final]{changes}

\newboolean{elsevier}
\setboolean{elsevier}{false}
\newboolean{sisc}
\setboolean{sisc}{false}
\newboolean{arxiv}
\setboolean{arxiv}{true}

\newboolean{notes}
\setboolean{notes}{false}

\usepackage{lipsum}                     % Dummytext
\usepackage{xargs}                      % Use more than one optional parameter in a new commands
\usepackage[colorinlistoftodos,prependcaption,textsize=tiny]{todonotes}
\newcommandx{\commentd}[2][1=]{\todo[linecolor=blue,backgroundcolor=blue!25,bordercolor=blue,#1]{#2}}
\newtheorem{assumption}{Assumption}

\title{Enclave Tasking for DG Methods on Dynamically Adaptive Meshes}

\ifthenelse{\boolean{elsevier}}{
  \input{paper-elsevier}
  
\begin{document}
  \begin{frontmatter}
 
  \begin{abstract}
  High-order Discontinuous Galerkin (DG) methods promise to be an excellent
discretisation paradigm for hyperbolic differential equation solvers running on 
supercomputers, since they combine high arithmetic intensity with localised data
access, since they straightforwardly translate into non-overlapping domain
decomposition, and since they facilitate dynamic adaptivity without the need for
conformal meshes.
An efficient parallel evaluation of DG's
weak formulation in an MPI+X setting however remains non-trivial as dependency
graphs over dynamically
adaptive meshes change with each mesh refinement or coarsening, 
as resolution transitions yield non-trivial data flow dependencies,
and as data sends along MPI domain boundaries have to be triggered in the correct
order.
Domain decomposition (MPI) alone starts to become insufficient if the mesh
changes very frequently, if mesh changes cannot be predicted, and if limiters
and nonlinear per-cell solves yield unpredictable cost per cell. 
We introduce enclave tasking as task invocation technique for shared memory and
MPI+X:
It does not assemble any task graph; instead the mesh traversal spawns 
ready tasks directly.
A marker-and-cell approach ensures that tasks feeding into MPI or
triggering mesh modifications as well as latency-sensitive or
bandwidth-demanding tasks are processed with high priority.
The remaining cell tasks form enclaves, i.e.~groups of tasks that can be processed in the background.
Enclave tasking introduces high concurrency which is homogeneously distributed
over the mesh traversal, it mixes memory-intensive volumetric DG calculations
with compute-bound Riemann solves, and it helps to overlap
communication with computations.
Our work focuses on ADER-DG and patch-based Finite Volumes.
Yet, we discuss how the paradigm can be generalised to the whole DG
family \added{and Finite Volume stand-alone solvers}.

  \end{abstract}

  \end{frontmatter}
}{}

\ifthenelse{\boolean{arxiv}}{
  \begin{document}

  \maketitle
 
  \begin{abstract}
  High-order Discontinuous Galerkin (DG) methods promise to be an excellent
discretisation paradigm for hyperbolic differential equation solvers running on 
supercomputers, since they combine high arithmetic intensity with localised data
access, since they straightforwardly translate into non-overlapping domain
decomposition, and since they facilitate dynamic adaptivity without the need for
conformal meshes.
An efficient parallel evaluation of DG's
weak formulation in an MPI+X setting however remains non-trivial as dependency
graphs over dynamically
adaptive meshes change with each mesh refinement or coarsening, 
as resolution transitions yield non-trivial data flow dependencies,
and as data sends along MPI domain boundaries have to be triggered in the correct
order.
Domain decomposition (MPI) alone starts to become insufficient if the mesh
changes very frequently, if mesh changes cannot be predicted, and if limiters
and nonlinear per-cell solves yield unpredictable cost per cell. 
We introduce enclave tasking as task invocation technique for shared memory and
MPI+X:
It does not assemble any task graph; instead the mesh traversal spawns 
ready tasks directly.
A marker-and-cell approach ensures that tasks feeding into MPI or
triggering mesh modifications as well as latency-sensitive or
bandwidth-demanding tasks are processed with high priority.
The remaining cell tasks form enclaves, i.e.~groups of tasks that can be processed in the background.
Enclave tasking introduces high concurrency which is homogeneously distributed
over the mesh traversal, it mixes memory-intensive volumetric DG calculations
with compute-bound Riemann solves, and it helps to overlap
communication with computations.
Our work focuses on ADER-DG and patch-based Finite Volumes.
Yet, we discuss how the paradigm can be generalised to the whole DG
family \added{and Finite Volume stand-alone solvers}.

  \end{abstract}
}{}

\ifthenelse{\boolean{sisc}}{

\author{
  D.E.~Charrier\footnotemark[2]
  \and
  B.~Hazelwood\footnotemark[2]
  \and 
  T.~Weinzierl\footnotemark[2]
}

\renewcommand{\thefootnote}{\fnsymbol{footnote}}
\renewcommand{\thefootnote}{\arabic{footnote}}
\slugger{sisc}{201x}{xx}{x}{x--x}%slugger should be set to mms, siap, sicomp, sicon, sidma, sima, simax, sinum, siopt, sisc, or sirev

\markboth{D.E.~Charrier and B.~Hazelwood and T.~Weinzierl}{Enclave tasking}

  \begin{document}
  \maketitle
  \begin{abstract}
  High-order Discontinuous Galerkin (DG) methods promise to be an excellent
discretisation paradigm for hyperbolic differential equation solvers running on 
supercomputers, since they combine high arithmetic intensity with localised data
access, since they straightforwardly translate into non-overlapping domain
decomposition, and since they facilitate dynamic adaptivity without the need for
conformal meshes.
An efficient parallel evaluation of DG's
weak formulation in an MPI+X setting however remains non-trivial as dependency
graphs over dynamically
adaptive meshes change with each mesh refinement or coarsening, 
as resolution transitions yield non-trivial data flow dependencies,
and as data sends along MPI domain boundaries have to be triggered in the correct
order.
Domain decomposition (MPI) alone starts to become insufficient if the mesh
changes very frequently, if mesh changes cannot be predicted, and if limiters
and nonlinear per-cell solves yield unpredictable cost per cell. 
We introduce enclave tasking as task invocation technique for shared memory and
MPI+X:
It does not assemble any task graph; instead the mesh traversal spawns 
ready tasks directly.
A marker-and-cell approach ensures that tasks feeding into MPI or
triggering mesh modifications as well as latency-sensitive or
bandwidth-demanding tasks are processed with high priority.
The remaining cell tasks form enclaves, i.e.~groups of tasks that can be processed in the background.
Enclave tasking introduces high concurrency which is homogeneously distributed
over the mesh traversal, it mixes memory-intensive volumetric DG calculations
with compute-bound Riemann solves, and it helps to overlap
communication with computations.
Our work focuses on ADER-DG and patch-based Finite Volumes.
Yet, we discuss how the paradigm can be generalised to the whole DG
family \added{and Finite Volume stand-alone solvers}.

  \end{abstract}

 
  \footnotetext[2]{
    Department of Computer Science, Durham University, 
    Lower Mountjoy South Road, Durham DH1 3LE, United Kingdom 
    (\email{tobias.weinzierl@durham.ac.uk})
  }
}{}

\section{Introduction}

%
% Context: why is DG a prominent technique
% - AMR
% - Arithmetic intensity
%
Higher-order Discontinuous Galerkin (DG) techniques contribute towards
the success story of many solvers for \added{hyperbolic} partial differential
equations (PDEs) on supercomputers.
DG methods are considered to be guarantors for computational efficiency.
While they fit to dynamically adaptive (block-structured) grids
\cite{Dubey:16:SAMR}---no conformity constraints are imposed
conceptually---DG's HPC selling point is that they combine high arithmetic intensity with
localised data access.
Its computations per mesh cell are arithmetically intense,
\added{
 which is a property they share with many higher-order methods
 \cite{Komatitsch:2010:HighOrder}.
 At the same time, DG's data access pattern however is very localised
 \cite{Dosopoulos:2011:MPIAndGPU}---this helps to reduce the memory access
 stress
 \cite{Charrier:19:RSC,Godel:2010:Scalability,Hindenlang:12:ExplicitDGForUnsteadyProblems,Klockner:09:DGonGPUs,Kronbichler:2017:FastDGKernels}---and
 its } exchange between cells along their connecting faces is conceptually simple.
A combination of these two properties---high intensity to exploit vector
units and dynamic AMR to invest where it pays off most---is a fit to predictions
what exascale software will have to look like
\cite{Dongarra:14:ApplMathExascaleComputing}.

%
% Meta-challenge
% - Why is shared challenging
% - What is a task graph/where does it arise?
% - Extremely heterogeneous task graph
% - AMR yields complicated dependencies hard to predict
% - Hide MPI? Difficult as too cheap
% - AMR and limiter render it even more complicated
% - Assembly not an option
% - We offer approach -> valid for all matrix-free DG
%
DG's localised data exchange is a fit to distributed memory, message-based (MPI)
parallelisation as it is predominant in supercomputing.
Delivering scaling algorithm
implementations per node as well as upscaling MPI+X yet remain far from trivial.
DG traverses the grid to evaluate its algorithmic steps.
Such traversals can be read as task graph traversals: 
The computational grid spans the graph whereas the particular DG scheme defines
the task type per mesh entity.
DG in its basic form distinguishes two tasks:
tasks working on cells and tasks working on faces (Riemann solves)
\added{\cite{Charrier:18:AderDG,Klockner:09:DGonGPUs}}.
Mapping the task types
onto separate mesh traversals makes steps piping data through the cores
(Riemann) take turns with the computationally demanding volumetric
evaluations.
While the latter scale, the other steps tend to be bandwidth- or latency-bound.
Furthermore, they couple cells and thus demand for data exchange.
They are thus typically incapable to exploit all cores, hinge on interconnect
capabilities, and their scalability potential is limited.
High polynomial degrees and static adaptivity with a volumetric coupling of
cells allow the cost for the cell operations to marginalise the cost for data
exchange, Riemann solves and so forth \cite{Charrier:18:AderDG,Heinecke:2014:PetascaleDynamicRupture}.
Overlapping domain decomposition with volumetric coupling is not studied here.
Nevertheless, cheap task phases continue to introduce a low concurrency
workload fraction in an Amdahl sense and thus constrain the scalability.

\added{
 Most codes that achieve high performance focus on particular DG subcategories
 and master the challenges above by exploiting the subcategories'
 particular characteristics:
 If we study classic Finite Volume (FV) schemes, the Riemann solve and the
 volume integral can be run in parallel
 \cite{Godel:2010:Scalability,Klockner:09:DGonGPUs}, i.e.~we can hide the former
 behind the expensive volumetric computation.
 If we study linear PDEs, the cost per cell is known a priori, as all cell and
 face operators are small matrices
 \cite{Heinecke:2014:PetascaleDynamicRupture,Klockner:09:DGonGPUs,Uphoff:17:SeisSol}.
 This simplifies the decomposition and scheduling of operations
 \cite{Burstedde:09:ALPS}.
 From a task point of view, static on-node scheduling then is sufficient.
 If we study static adaptive meshes, we know prior to each time
 step where computationally intense interpolations and restrictions arise that
 feed into other tasks.
 We in particular know which face data are to be exchanged via MPI which enables
 us to prioritise the handling of the underlying computations appropriately such that they are sent out while we
 continue to work locally
 \cite{Ferreira:2017:LoadBalancingAndPatchBased,Godel:2010:Scalability,Heinecke:2014:PetascaleDynamicRupture,Hindenlang:12:ExplicitDGForUnsteadyProblems,Komatitsch:2010:HighOrder,Mu:2013:Accelerating}.
 We can
 even design hardware topology-aware domain splits
 \cite{Sundar:15:Enclave}.
 If we furthermore stick to conformal
 meshes plus global time stepping or even regular grids, such operations
 disappear completely
 \cite{Dumbser:18:EfficientADERDG,Heinecke:2014:PetascaleDynamicRupture,Uphoff:17:SeisSol}.
 These success stories show that DG has great upscaling potential.
 They also show that it is 
}
reasonable to reorder and intermix the tasks
\added{
 to obtain high performance.
}

%
% Gilt das nur fuer dynamisches AMR?
%

%
% Technical details of challenge
%
Any rearrangement or parallelisation of the task execution requires
care as inter-grid transfer operations along mesh resolution boundaries 
have to be performed in the correct order.
We typically interpolate the coarser data representation,
solve Riemann problems then, and finally restrict the outcome.
Furthermore, tasks sending and receiving MPI messages have to stick to a
specific order.
The fact that there are ``cheap'' tasks, i.e.~tasks with low arithmetic
intensity, further implies that MPI sends have to
be issued early to allow the message transfer to hide behind the expensive
tasks or many cheap ones. 
Finally, memory-intense tasks such as mesh refinement or the Riemann solves
shall continuously trickle through the system to avoid memory access bursts.
Assembling the whole task graph or fractions of it and then deriving a
tailored/optimised schedule without assembly penalty is difficult 
if totally dynamic, unconstrained
AMR may change the graph in each and every time step and (almost) any location
in the computational domain.
\added{
 The above paragraph gives examples of successful strategies if we constrain the
 AMR.
}
Some sophisticated DG variants furthermore employ techniques such as 
(a posteriori) limiting \cite{Dumbser:14:Posteriori}, optimistic time stepping
with on-the-fly CFL analysis which occasionally require roll backs, or solve
nonlinear equation systems locally with dynamic termination criterion
controlling the nonlinear solve.
The cost per volume is not known a priori.
Local time stepping is beyond scope in the present work yet can be seen as
technique which amplifies all task balancing and scheduling difficulties.

%
% What do we do and how does it compare to literature
%
We propose a grid traversal and task invocation scheme called {\em
enclave tasking}.
It works without any task graph assembly,  and it makes no assumption about the
grid topology.
We introduce it by means of dynamically adaptive meshes as they result
from octrees as well as generalisations\added{---we call them spacetrees---}of
those \cite{Weinzierl:19:Peano,Weinzierl:11:Peano},
and by means of ADER-DG
\cite{Dumbser:18:EfficientADERDG}, an explicit time stepping scheme for hyperbolic equation systems.
The spacetree is traversed \added{cell-wisely}.
This allows for many efficient storage and traversal schemes
\cite{Dubey:16:SAMR,Griebel:98:HashStorage,Weinzierl:11:Peano,Weinzierl:19:Peano}.
Enclave tasking maps each DG time step onto a pair of mesh traversals.
The primary mesh traversal runs over all mesh cells and spawns one
task per cell.
Work stealing then distributes these computationally intense tasks
among idle threads.
We realise a producer-consumer pattern.
Additionally, the traversal launches the computationally cheap
Riemann solves ad hoc, i.e.~per face read:
it waits for the tasks of the adjacent cells to
terminate, and then it immediately runs the bandwidth-demanding computations.
Where required, the primary mesh traversal processes a temporarily shifted task
graph \cite{Charrier:18:AderDG}:
It runs the Riemann solves, and then immediately issues the
cell tasks of the subsequent time step.
% If global time stepping with prescribed time step sizes is used, the correction
% can be merged into the subsequent prediction task.
%%
%
Cell tasks update the solution within their cell,
which implies a change of the
solution representation along the cell faces.  
In our non-overlapping domain decomposition,
these face data have to be sent out in a deterministic, consistent
order to neighbouring ranks, as
the subsequent primary mesh traversal on the neighbouring rank feeds
its local data plus the remote counterpart data as input to the Riemann solves.
We thus classify the aforementioned cell tasks into high priority and background
tasks, and we introduce a secondary (partial) mesh traversal.
It takes turns with the primary traversal.
Whenever the secondary traversal accesses a cell along a domain boundary, 
it waits for its local adjacent task to complete and then passes the
outcome immediately to MPI.
Tasks of cells adjacent to MPI boundaries are issued with high priority.
Furthermore, we assume that dynamic adaptivity spreads along existing grid
transitions most of the time. 
It tends to evolve smoothly in space and time.
Therefore, we also make cell tasks along refinement transitions have high
priority if they feed into a mesh interpolation. 
Their outcome is processed by the secondary mesh traversal, too.
This means that
information from interpolation along resolution boundaries becomes
available early.
Other cells are skipped by the secondary traversal.
As our mesh traversals itself are parallelised, too, we end up with three
different types of tasks spread over two different types of mesh traversals:
Memory-intense tasks tied to the (parallel) mesh traversals,
high priority cell tasks and background cell tasks.
The high priority tasks stem from cells adjacent to MPI boundaries and
adaptivity.
They form {\em skeletons}.
The remainder cells form {\em enclaves}.
They are tasks that are handled
in the background of all communication- and bandwidth-critical operations.
They deliver the scalability.

%
% Our contribution/work and comparison
%
The whole paper describes a geometrically inspired multitasking scheme which exploits
mesh regularity and the fact that meshes typically do not change 
dramatically all over the domain within a short time interval.
The idea to process ``communicating cells'' prior to others
is, notably in the context of DG and accelerators, not new
\cite{Baggag:1999:Parallelization,Hindenlang:12:ExplicitDGForUnsteadyProblems,Komatitsch:2010:HighOrder,Sundar:15:Enclave},
\added{and many of the present ideas, per se, are well-known.}
To the best of our knowledge, there's however no work
\added{that combines all of the following features into one formalism. Our work} 
(i) derives and updates regularity information---the
skeletons---on-the-fly, and thus imposes no constraints on the dynamic
adaptivity.
In our code, the adaptivity along subdomain boundaries can change in each
and every grid sweep;
(ii) does not make any assumptions about the grid structure/topology or
restricts itself to particular subgrid regions/enclaves
\cite{Komatitsch:2010:HighOrder,Schreiber:13:Clusters,Sundar:15:Enclave}; 
(iii) works without any assembly of a task graph which becomes expensive if
dynamic adaptivity makes this graph change in each and every grid sweep
\cite{Kormann:2011:Parallel,Kronbichler:2017:FastDGKernels,Schreiber:13:Clusters}
\added{and yet supports very inhomogeneous, unpredictable cost-per-cell
distributions}; (iv) mixes tasks with different compute
characteristics and thus avoids memory access bursts.
(v) Enclave tasking finally is MPI-oblivious:
As MPI data aligns along the mesh skeleton, MPI data is sent out while
prediction tasks still might queue in the background.
Therefore, sends are overlapped with computation.
Notably, enclave tasking does not have to know the MPI communication pattern a
priori, which would be a showstopper for totally free dynamic adaptivity.
% Our code even scales in the presence
% of dynamic adaptivity plus a posteriori limiting.
% To the best of our knowledge, no other ADER-DG implementation has combined these
% characteristics before.
While our discussion focuses on ADER-DG, a particular flavour of explicit time
stepping DG schemes, we sketch that our techniques
impact many matrix-free DG methods and notably all
Finite Volume schemes, too.
\added{
While our discussion focuses on spacetrees, the extensions to forests
\cite{Bangerth:11:dealiiwithp4est,Isaac:15:RecursiveAlgorithmsOnOctrees,Sasidharan:16:MiniAMR}
is straightforward if we classify all tree, i.e.~inter-forest boundaries as
skeletons.
}

%
% Organisation
%
The remainder is organised as follows:
We sketch ADER-DG and the operators, i.e.~tasks of interest
(Sect.~\ref{section:ader-dg}), before we introduce enclave
tasking in Sect.~\ref{section:enclave-tasking}.
Section \ref{section:tasking} next describes
how we tailor the tasking runtime and use MPI.
In Sect.~\ref{section:dg}, we generalise all patterns to other
DG approaches and Finite Volumes (FV), and then provide 
measurements (Sect.~\ref{section:results}) that demonstrate the potential of
our ideas. 
A conclusion summarises the main findings and sketches future work as well as
shortcomings.

\section{ADER-DG on Cartesian meshes}
\label{section:ader-dg}
 
We study first-order hyperbolic systems 
% which can be written as
\begin{equation}
  \frac{\partial Q}{\partial t}
 +
 \nabla\cdot F(Q)
%  +
%    \sum_{i=1}^d B_i(Q) \frac{\partial Q}{\partial_{x_i}}
 = 
   S(Q)
 +
    \sum \delta
    \qquad
    \mbox{with}\ 
    Q: \mathbb{R}^{d+1} \mapsto  \mathbb{R}^m,
    d \in \{2,3\}.
 \label{equation:introduction:PDE}
\end{equation}
$F$ is a conservative flux, 
$S$ a volumetric source term, and $\delta $ 
denotes the impact of point sources.
$S$ and $\delta $ usually depend on time
\added{and space.}
$Q$ is the solution over a $d$-dimensional computational domain.
It has $m$ components and changes over time.
The system is complemented by well-suited initial and boundary conditions.

%
% Specialisation: ADER-DG
%
% - Higher order
% - CFL
% - High order in time/explicit plus stability
Among DG techniques for (\ref{equation:introduction:PDE}), ADER-DG
\cite{Dumbser:06:ADERDG,Dumbser:14:Posteriori} has grown into a popular
approach.
ADER-DG relies on particularly expensive volumetric cell operators,
as it solves the PDE per mesh cell per time step through a weak space-time
formulation.
For nonlinear PDEs, this even requires a nonlinear equation system solve.
Space-time solves are 
computationally feasible as the mesh cells are handled independently of
each other.
The solve, however, is only a \added{space-time predictor} which
feeds into a follow-up, explicit-in-time Riemann phase.
It solves the Riemann problems arising from the discontinuous, predicted
solution along the mesh faces.
Eventually, both solve outcomes are merged into the next time
step's solution.
This step is labelled corrector.
ADER-DG exhibits high-order behaviour in both space and time, and it is
arithmetically intense per cell, 
although it requires only one data exchange between adjacent cells per time
step.

\subsection{ADER-DG sketch} 

We study (\ref{equation:introduction:PDE}) over a
computational domain $\Omega _h$ discretised by a mesh that 
consists of cuboid cells $c$.
% \commentd{Schlussendlich setzen wir dann aber $Q_h(x,T) = func(x)$ in die Zellen.}
Each cell carries an $Q$ approximation $Q_h(x,t)$ as linear combination of 
Lagrangian polynomial shape functions
over 
% Gau\ss-Lobatto or 
Gau\ss-Legendre points.
The polynomials are continuous inside the cells, but they induce jumps along
the faces between cells.
Our derivation of the algorithmic steps and computational tasks---each step
consists of tasks which are atomic work units independent
of all other tasks---derives 
from \cite{Dumbser:18:EfficientADERDG}.

ADER-DG starts from a weak formulation of (\ref{equation:introduction:PDE}) both
in space and in time.
%  over the time interval $(T,T+\Delta T)$.
Let $(T,T+\Delta T)$ span one time step. 
We obtain a continuous, weak formulation
\begin{equation}
 \int _{\Omega \times(T,T+\Delta T)}
 \Bigg( 
 \frac{\partial Q}{\partial t}
  +
 \nabla\cdot F(Q)
 \Bigg)
 \hat{v}\,
 \mathrm{d}(x,t)
  = 
\int _{\Omega \times(T,T+\Delta T)}
\Bigg(
   S(Q)
  +
    \ldots
 \,
\Bigg)
\hat{v}\,
 \mathrm{d}(x,t)
 \label{equation:aderdg:weak-space-time}
\end{equation}

\noindent
where we replace $Q(\cdot, T)$ by the linear combination of shape functions
 $Q_h$, and
where we vary a space-time test polynomial $\hat{v}$.
% \commentd{spatial test polynomial $\hat{v}$ im corrector, space-time nur im
% Predictor.}
To develop $Q_h$ in time, we multiply it with a polynomial in time
\cite{Dumbser:06:ADERDG}. 
The same Lagrangian polynomial order as for the spatial representation is used.
With cubes as mesh cells, this is a tensor product approach for the space-time
solution. 
It describes a space-time polynomial $\hat Q_h$ approximating the development
of the real solution $Q$ over time and space.
$\hat{v}$ uses the same ansatz, i.e.~we express ${Q}_h(x,t)$ or $\hat
{Q}_h(x,t)$, respectively, from (\ref{equation:aderdg:weak-space-time}) by a
weak, discretised space-time Ritz-Galerkin problem with space-time test functions $\hat v_h$.

Following \cite{Dumbser:18:EfficientADERDG}, we separate the time
derivative from the remainder integrals in 
(\ref{equation:aderdg:weak-space-time}).
Partial integration in time for the term comprising $\frac{\partial Q}{\partial
t}$, 
and partial integration in space for all other terms gives us two
computational/implementational advantages.
(i) It injects the known solution $Q_h(\cdot,T)$ into the system and yields an
explicit expression for $Q_h(\cdot,T + \Delta T)$, as we roll over the
time derivative to the test function.
(ii) It removes the divergence operator $\nabla \cdot $ from $F$ as it
\added{transfers}
to the test function. 
$F$ evaluations for applications are straightforward---they typically
describe the physics directly---while a derivative computation can be tedious.
% The handling of non-conservative terms $B$ requires additional work
% and is beyond scope here.
The two advantages are accompanied by two disadvantages:
(i) The overall scheme is high order in time but describes a globally implicit
setup which is usually infeasible to solve.
(ii) We inherit jump terms from the partial integration in space.
ADER-DG addresses these two disadvantages numerically.

The first step of ADER-DG is the \emph{space-time predictor} (STP).
It develops $\hat {Q}_h(x,t)$ by solving the
weak, discretised space-time Ritz-Galerkin problem.
Yet, it drops all jump terms when it integrates

 \[
 \int _{c\times(T,T+\Delta T)}
 \Bigg(
 \frac{\partial Q_h}{\partial t}
 \Bigg)
 \hat{v}\,
 \mathrm{d}(x,t)
  +
 \int _{c\times(T,T+\Delta T)}
 \Bigg(
 \nabla\cdot F(Q_h)
 \Bigg)
 \hat{v}\,
 \mathrm{d}(x,t)
  = 
\ldots
 \]

\noindent
by parts.
With $\bigcup c = \Omega _h $ tessellating the computational domain, the arising
STP decouples the individual cells from each other.
We ignore the neighbours of any cell $c$.
For a given $Q_h(\cdot,T)$, this yields a space-time $\hat Q_h^* $
through Picard iterations.
%  unless (\ref{equation:introduction:PDE}) is linear.
% In the linear case, Cauchy-Kowalevski yields a linear equation system
% \cite{Dumbser:06:ADERDG,Dumbser:14:Posteriori,Dumbser:18:EfficientADERDG}.

As $Q_h,\hat Q_h,\hat Q_h^*$ all are represented by polynomials with compact
support that are allowed to be discontinuous along cell faces,
no continuity constraints between cells are built into $\hat Q_h^*$.
Jumps arise if we extrapolate the STP $\hat Q_h^*$
from left and right to the faces between neighbouring cells.
Such projections are labelled $\hat Q_h^{*\pm}$ in DG---one 
projection from the right and one from the left along a coordinate axis.
As $\hat Q_h^{*\pm}$ is a space-time polynomial, the extrapolation is a
space-time expression, too.
To make ADER-DG an explicit time stepping scheme,
we replace all $\hat Q_h$ entries in the
``jump'' terms arising from 
partial spatial integration of (\ref{equation:aderdg:weak-space-time}) with our
predicted $\hat Q_h^*$.
% The STP $\hat Q_h^*$ is used to construct the interface states between any
% two cells on $(T,T+\Delta T)$.
These face interface states then are plugged into a Riemann solver. 
This is the second step of the ADER-DG scheme.
We solve the {\em Riemann} problems.

In the third and final step of ADER-DG, we plug both $\hat
Q_h^*(\cdot,T+\Delta T)$ and the time integral over the Riemann solution into
(\ref{equation:aderdg:weak-space-time}), integrate in time and solve the
remaining weak formulation.
It degenerates to a spatial problem over $\Omega _h$.
This step can be read as a correction to the predicted $\hat
Q_h^*(\cdot,T+\Delta T)$.
It is thus called {\em corrector}.

\subsection{A task language}
Let $C_\text{STP}$ denote 
\begin{equation}
\hat Q_h^* = C_\text{STP}\,Q_h(\cdot,T).
\label{equation:ader-dg:C}
\end{equation}
\noindent
In (\ref{equation:ader-dg:C}), we use the symbol $C_\text{STP}$ as global operator
applied to the solution over the whole computational domain.
However, the STP's construction implies that $C_\text{STP}$
decomposes over the cells.
Consequently, we use $C_\text{STP}$ synonymously for a computational task which
advances the solution over one cell: 
$\hat Q _h^* | _{c} = C_\text{STP} \cdot Q(\cdot,T)  | _{c}$.
We omit $ | _{c} $ from hereon.
As a result,
the global $C_\text{STP}$ evaluation results from the application of a set of
cell-wise $C_\text{STP}$ tasks to all $c \in \Omega _h$.
Though $C_\text{STP}$ formally spawns the whole STP, follow-up steps
use solely $Q_h^*(\cdot,T+\Delta T) = id|_{T+\Delta T} \cdot \hat Q_h^*$ and the
two projections $id |_{\partial c} \cdot \hat Q_h^*$ of the STP onto each
face between any two cells.
% TODO
It is convenient only to store these results
\cite{Charrier:18:AderDG}, i.e.~to extrapolate---if we employ
Gauss-Legendre points, no sample points coincide directly with the faces---and
to integrate over time immediately.

Let $F_\text{R}$ denote the operator that runs over all faces.
It represents the Riemann solves. 
For global time stepping, where all cells advance in time with the same time
step size $\Delta T$, it is convenient to make it comprise the time integral
over the result, too.
Along the lines of (\ref{equation:ader-dg:C}), 
% we use $F_\text{R}$, on the one hand, to
% denote the global update from a predicted solution into the next time step's solution.
% On the other hand, 
we observe that $F_\text{R}$ decomposes over the mesh faces.
$F_\text{R}$ consequently describes a set of compute tasks over all faces.
They accept input from the $C_\text{STP}$ tasks.

The {\em corrector} finally yields a set of cell-wise $C_\text{Corr}$ tasks.
We end up with
\begin{eqnarray*}
  Q_h(\cdot,T+\Delta T) & = & 
   C_\text{Corr} \circ F_\text{R} \circ  id |_{\partial c} \cdot
   Q_h^* + id|_{T+\Delta T} \cdot Q_h^* \nonumber \\
  & = & \left(
   C_\text{Corr} \circ F_\text{R} \circ  id |_{\partial c} 
   + id|_{T+\Delta T}
   \right)  \circ C_\text{STP} \cdot Q_h(\cdot,T)
\end{eqnarray*}

\noindent
for one ADER-DG time step.
Alternatively, we may distinguish the data
stored inside the cell from the data held for the faces by writing them as
entries of a vector:

\begin{eqnarray}
 \left( \begin{matrix} Q_h \\ \cdot \end{matrix} \right)(T+\Delta T)
 &=&  
 \underbrace{
 \left( \begin{matrix} 1 & C_\text{Corr} \\ 0 & 0 \end{matrix} \right)
 }
 _{\mbox{\footnotesize corrector (cell-wise)}}
 \underbrace{
 \left( \begin{matrix} 1 & 0 \\ 0 & F_\text{R} \end{matrix}
 \right) 
 }
 _{\mbox{\footnotesize Riemann (face-wise)}} 
 \nonumber
 \\
 &&
 \underbrace{
 \left( \begin{matrix} id|_{T+\Delta T} \\ id|_{\partial c} \end{matrix}
 \right) 
 C_\text{STP}
 \left( \begin{matrix} 1 & 0 \end{matrix} \right)
 }
 _{\mbox{\footnotesize STP (cell-wise)}}
 \left( \begin{matrix} Q_h \\ \cdot \end{matrix} \right)(T).
 \label{equation:ader-dg:tasks}
\end{eqnarray}

\noindent
Face data here is used as temporary data storage and thus does not
determine the solution at particular time stamps.
Once we are given a mesh,
(\ref{equation:ader-dg:tasks}) describes the arising ADER-DG task graph.
Alternatively, 
this ADER-DG blueprint maps onto a plain realisation employing three loops (Algorithm~\ref{algorithm:ader-dg-pseudo-code})
each issuing independent tasks.

%\vspace{-0.2cm}
\begin{algorithm}[htb]
  \caption{
    Pseudo-code of ADER-DG split up into three phases. We highlight what we
    refer to as $C$ and $F$ tasks. Some technical details (projections and
    temporary variables) from this algorithm are omitted in the text for
    clarity.
    Without local time stepping, time integration and Riemann solve $F_\text{R}$
    can be switched, i.e.~we can collapse Riemann solves over time into one
    spatial Riemann problem.
  }
  \label{algorithm:ader-dg-pseudo-code}
  \begin{algorithmic}[1]
    \Function{AderDGTimeStep}{$\Delta T$} 
    %\DontPrintSemicolon \;
    \ForAll{cells $c \in \Omega _h$}
      \Comment Space-time predictor phase
      \State \Call{startTask}{}
      \State \Call{$C_\text{STP}$}{$c$}
        \Comment Run predictor on cell. $C_\text{STP}$ is parameterised with $\Delta T$ 
      \State \Call{$id_{\partial c}$}{$c$}
        \Comment Project result to $2d$ faces of $c$, keep outcome at
        $T + \Delta T $, 
      \State
        \Comment too, but throw away intermediate time solutions
      \State \Call{endTask}{}
    \EndFor
    \ForAll{faces $f \in \Omega _h$}
      \Comment Riemann phase
      \State \Call{startTask}{}
      \State Read $id|_{\partial c}$ outcome from $f$'s adjacent cells
      \Comment $Q_h^{*\pm}$ in DG literature
      \State $F_\text{R}(f)$
        \Comment Riemann solves over the whole $(T,T + \Delta T)$ time span
      \State Integrate Riemann outcome in time
      \State \Call{endTask}{}
    \EndFor
    \ForAll{cells $c \in \Omega _h$}
      \Comment Correction phase
      \State \Call{startTask}{}
      \State Read $F_\text{R}$ outcome of $2d$ adjacent cells
        \Comment Project Riemann solve 
      \State $C_\text{Corr}(c)$ 
        \Comment outcomes back to cell, and fuse with predicted data
      \State \Call{endTask}{}
    \EndFor
    \EndFunction
  \end{algorithmic}
\end{algorithm}
%\vspace{-0.2cm}

\subsection{Dynamic adaptivity}

While adaptive mesh refinement (AMR) minimises computational work, it adds complexity to the task graph.
For ADER-DG, we are however able to exploit the discontinuities build inherently
into the \added{numerical scheme} to bring down AMR implementation complexity
from a parallelisation point of view.
We do not impose any balancing conditions
\cite{Sundar:08:BalancedOctrees}.
Yet, we do assume that we have a grid topology which is common to many software packages:

\begin{assumption}
Let our grids result from a conformal grid.
We assign this grid the level $\ell _{min} $.
From $\ell _{min} $ on, we construct a finer, adaptive grid by
subdividing each cell that we want to refine a fixed number of times
such that the subdivisions along two adjacent cells that are refined match.
This new grid has level $\ell _{min}+1 $.
We continue recursively but independently for each cell.
When the recursion has terminated, all cells that are not subdivided further
form an adaptive mesh $\Omega _h$.
Each cell $c \in \Omega _h$ belong to a unique level.
\end{assumption}

\noindent
The purest grids of this type are quadtrees and octrees.
They start from one square or cube, i.e.~a trivial $\ell _{min}$ mesh, and
subdivide this base cell along each coordinate axis once per refinement step.
They eventually yield an adaptive Cartesian mesh where all cuboids of one
level have exactly the same size.
Our grid assumption includes forest of trees where we start from a
conformal mesh and embed octrees into its cells.
The extension to more sophisticated subdivision or boundary-fitted meshes is  
straightforward.
The hyperbolic ecosystem around \cite{Berger:89:AMR}
imposes a grid topology suiting our assumption, too.
For our experiments, we stick to sole trees.
Different to the traditional bipartitioning, we use three-partitioning
\cite{Weinzierl:19:Peano,Weinzierl:11:Peano}.

Along each resolution transition of the resulting fine grid $\Omega _h$, we can
uniquely identify cells of the coarse and the fine resolution.
AMR now becomes a strict extension not
altering any building blocks introduced so far:
As the STP and the correction are tied to cells, they are
agnostic of AMR.
For the Riemann problems, we first introduce \emph{virtual
cells}.
Virtual cells are subcells of real cells which have the same size, i.e.~face
lengths, as their adjacent real cells.
No virtual cell overlaps two real cells by construction.
To obtain the Riemann preimage for the child face, we
interpolate $\hat Q_h^*$ from the coarser cell into its virtual cells.
The virtual cell then extrapolates 
its ``inherited'' STP through $id|_{\partial c}$
onto the face, where it complements the data from the real cell.
Interpolation is realised from coarse to fine cell resolutions,
prior to the respective Riemann solves.
\added{
The concept to use virtual cells to remove hanging faces temporarily, i.e.~from
a solver's point of view, is well-established in the (block-)structured  world,
where it is also known under the term ghost cells
\cite{Berger:98:AMR,LeVeque:11:TsunamisAMR} or halo layer
\cite{Sasidharan:16:MiniAMR}.
}

The Riemann outcome along the interface of a coarse and fine cell
affects the coarse cell's corrector.
Here, we switch from a compute-Riemann to an
accumulate-Riemann approach.
Let a face along a resolution transition be a parent face it belongs to the
coarser fine grid level.
Let the term child face refer to a segment of this parent face which coincides
with the face of one adjacent finer cell.
Where a face is parent to other faces, no 
Riemann solve is applied to the parent, i.e.~the coarse face.
Instead, 
Riemann solves along resolution transitions are always computed along their
finest resolution.
Every time we determine the Riemann outcome along a child face, we accumulate 
it back into the parent face.
This is a restriction (the transpose of the interpolation).
For it, we traverse the adaptive mesh starting from
the finest cells.

Adaptive meshes require us to introduce two additional tasks to
(\ref{equation:ader-dg:tasks}):
Interpolation and restriction.
The volumetric interpolation of $\hat Q_h^*$ can be subject of
optimisation---only the projections onto the virtual cells' faces are required.
The important observation however is that the interpolation is a pure
preprocessing step to the Riemann solves that squeeze in-between $C_\text{STP}$ and
$F_\text{R}$.
Interpolation incorporates resolution logic, but it does not impose
additional partial order constraints on either the STP or the Riemann
solves.
An analogous observation is to be made
for the restriction of the Riemann solves' outcome.

Dynamic adaptivity adds further tasks.
We focus on feature-based refinement criteria and assume that codes decide
cell-wisely whether to refine or coarsen.
They read the solution $Q_h(\cdot,T+\Delta T)$ and study the solution's
character.
They thus introduce an epilogue to $C_\text{Corr}$.
If the criterion triggers a refinement, new cells are created, i.e.~volumetric
data are interpolated, and the algorithm updates the virtual cells.
This has to complete prior to any STP task in the affected part of the
computational domain.
If the coarsen criterion identifies a cluster of cells which can be coarsened
into one bigger cell, we trigger an analogous workflow.
Since the analysis of the cell-wise refinement/coarsening criterion is strictly
element-wise, all tasks thus can run in parallel.
The update of the virtual cells and the merger of small cells into a bigger one
induce causal dependencies which however are localised and simple to integrate
into the task graph.

\subsection{Computational character}

 %
 % Arithmetic intensity
 %
As both correction and Riemann tasks in (\ref{equation:ader-dg:tasks})
rely on the same PDE terms---through the partial integration on
(\ref{equation:aderdg:weak-space-time}) the lion's share of the compute load
results from $F$ evaluations in (\ref{equation:introduction:PDE})---their
abstract arithmetic intensity \cite{Williams:09:Roofline}, i.e.~their sole computations-per-double ratio, is
comparable.
This statement holds if the employed Riemann solver---we
use Rusanov here---only requires $F$ and few additional data such as an estimate
of the biggest eigenvalue.

$C_\text{STP}$ however integrates over polynomials in space and time.
They are typically stored in small continuous array blocks per cell.
If the STP is an iterative solve, this solve plus the time 
integration lead to a high intensity relative to
the caches \cite{Ilic:14:CacheAwareRoofline}:
The arithmetic intensity given as ratio of operations to loads from the main
memory into the registers or a reasonably close cache is expected to be
relatively high.
If we study a linear variant of (\ref{equation:introduction:PDE}), we 
integrate the cell with the Cauchy-Kowalesvki procedure \cite{Dumbser:06:ADERDG}.
Here, the STP is significantly cheaper though it still yields localised data access \cite{Charrier:19:RSC}.
The time integration following the STP allows us to
reuse the outcome data structure for all intermediate-in-time results.

In contrast, the Riemann solves are cheap and explicit.
With the volumetric terms in (\ref{equation:introduction:PDE}), i.e.~sources and  
point terms, disappearing, an Riemann tasks loads in the
predicted solution and writes back its result to the respective face.
The corrector finally is of similar simplicity, as it takes two input data
streams and yields $Q_h(T+\Delta T)$.
The remaining volumetric integration is simplistic.
Also interpolation and restriction tasks are neither sophisticated nor
computationally demanding as they are based upon the polynomials.

Finally, all refinement criteria we encounter in our code base are conceptually
simple.
We study the first or second order derivatives of the solution and make our
refinement or coarsening decision from there.
With an explicit polynomial representation of $Q_h$ being available, these tasks
therefore are of low computational intensity.
They however have to read the whole volumetric data, and the subsequent
refinement or coarsening might induce further memory accesses and allocations.

\begin{assumption}
 We assume that the  STP, i.e.~the volumetric task is
 computationally heavier than all other tasks.
 In particular the face task (the Riemann solve) tends to be memory
 bandwidth- and latency-bound.
 It brings together data from adjacent cells scattered in memory but does not
 yield a high number of floating point operations.
\end{assumption}

\begin{figure}[htb]
  \begin{center} 
   \includegraphics[width=0.42\textwidth]{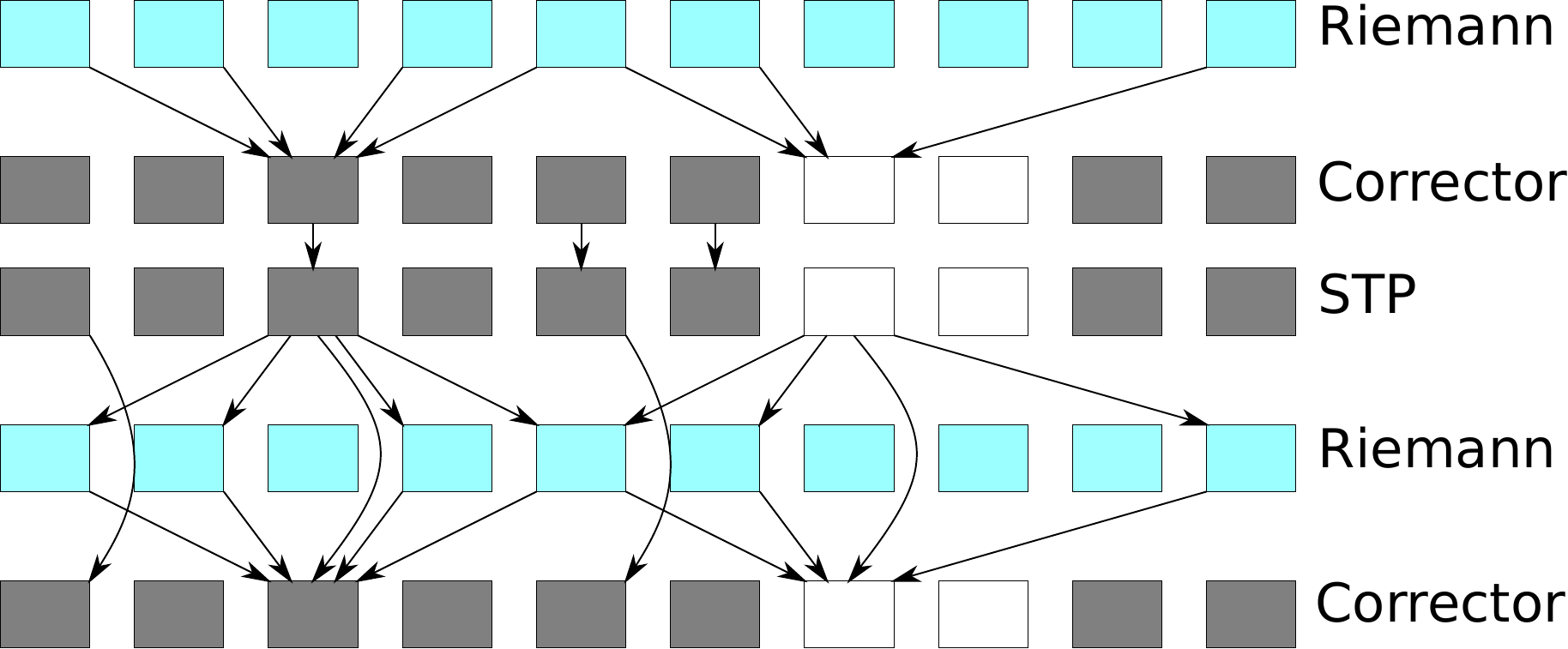}
   \hspace{0.02\textwidth}
   \includegraphics[width=0.2\textwidth]{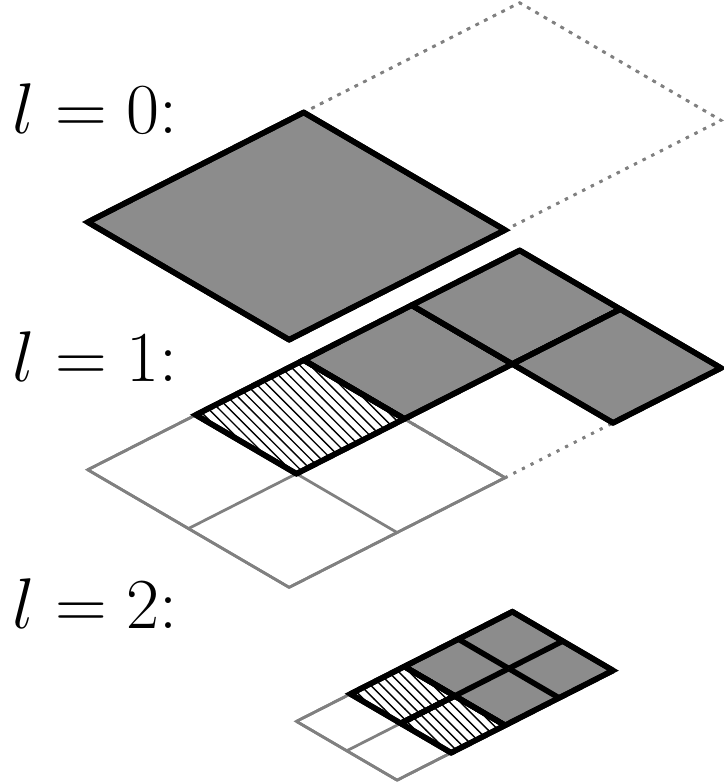}
   \hspace{0.02\textwidth}
   \includegraphics[width=0.30\textwidth]{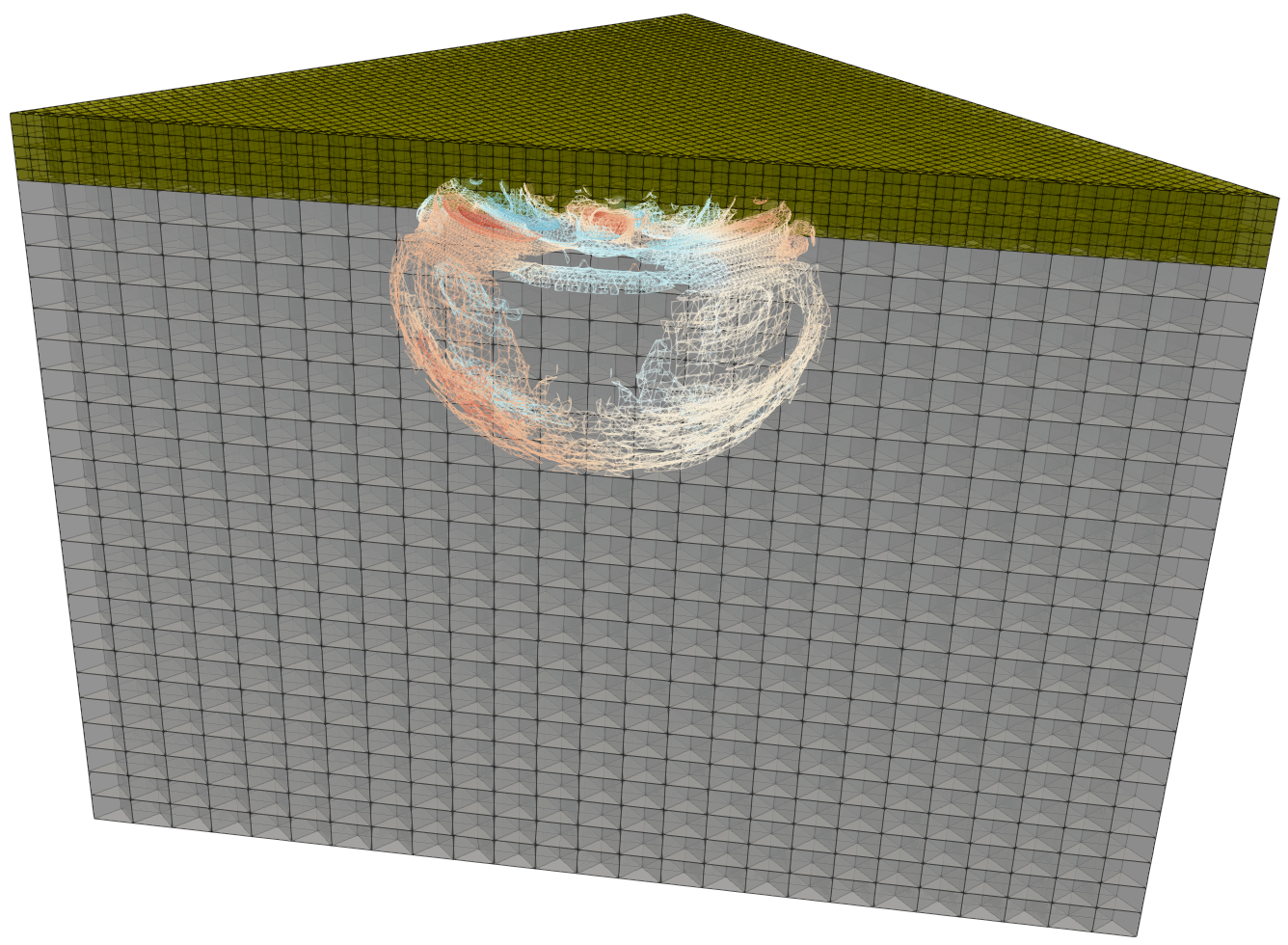}
  \end{center}
 \vspace{-0.2cm}
  \caption{
    Left: Schematic illustration of task graph with some dependencies for
    two-dimensional setup:
    Cells within a regular grid region (filled) combine the local DG solution
    with the result of $2d$ Riemann solves.
    Cells adjacent to grid refinement (empty) require input from more
    or less Riemann solves if they run the corrector.     
    Middle: Schematic illustration of various spacetree (quadtree) nodes when a
    tree is used to host an enclave/skeleton mesh.
    Dark tree nodes are fine grid nodes. 
    They can be refined ($\ell =0$) or unrefined ($\ell \in \{1,2\}$).
    Nodes with hatching are (unrefined) virtual nodes. 
    White nodes are supplemental. 
    They complete the tree but they do not carry data.
    Right: Sketch of the LOH.1 benchmark: It is a simplified earthquake
    setup, where the cubic domain consists of two types of material (layers)
    and a point source stimulus induces the elastic wave propagation.
    \label{figure:enclave:tree}
  }
 \end{figure}
 
\subsection{Task graph structure}
ADER-DG's task graphs are conceptually simple
(Fig.~\ref{figure:enclave:tree}):
There are cell tasks, face tasks and AMR tasks.
For simplicity, 
we omit an explicit discussion of tasks tied to dynamic and static adaptivity
for the remainder of this section where appropriate.
We note that (i)
the STP decomposes into one task per cell.
(ii) The individual $C_\text{STP}$ tasks are independent of each other.
Operator $F_\text{R}$ decomposes into one task per grid face.
(iii) Finally, the individual $F_\text{R}$ tasks are independent of each other.
While each $F_\text{R}$ task requires input from the $C_\text{STP}$ tasks from its adjacent
cells,
each $C_\text{Corr}$ task requires input from the $2d$ $F_\text{R}$ tasks of the adjacent
faces.

Our task types translate into two
grid traversal types: One over cells, one over faces.
Per type, all task evaluations are
independent of each other.
Task assembly-free processing thus is possible if we run over the grid
three times.
A first traversal issues all $C_\text{STP}$ tasks and eventually waits for them
to complete before a second traversal issues all $F_\text{R}$ tasks. 
A final sweep corrects the solution and thus yields the subsequent time step's
solution.
Such an assembly-free approach describes a producer-consumer pattern:
One or few main threads traverse the grid and produce tasks, all
other threads consume these.

We assume that our grid changes frequently. A na\"ive realisation with
% \commentd{three grid sweeps?}
grid sweeps exhibits disadvantageous properties:
 \begin{enumerate}
   \item We employ a non-overlapping domain decomposition and solve the Riemann
   problems redundantly on both adjacent ranks.
   A STP adjacent to an MPI
   domain boundary thus has to send its face data over to
   neighbouring ranks, such that all ranks can run their
   $F_\text{R}$ tasks autonomously.
   Data exchange in MPI has to be deterministic. We may not simply spawn
   $C_\text{STP}$ tasks and make them send their outcome.
%    There are temporal dependencies.
   \item Modern multicore chips are equipped with memory controllers that cannot
   keep all cores busy.
   Algorithms have to avoid that all tasks access the main memory controllers
    concurrently and thus become
    bandwidth-bound
   \cite{McCalpin:95:Stream,Williams:09:Roofline}.
%    Notably, they have to avoid that too many bandwidth-intense tasks run at the
%    same time. 
   Fire-and-forget of $F_\text{R}$ tasks by the sweeps however yields
   a large set of memory-sensitive tasks in one rush.
   \item Cache-efficient codes perform as many operations as possible on
   data before these are moved out into the main memory again.
   %Even though we launch each $C_\text{STP}$ immediately after ``its'' corrector, 
   With one sweep per phase, we have to assume
   that the outcome of a Riemann solve does not reside inside the cache long
   enough for the next corrector.
   A similar consideration holds for the outcomes of the STP.
   \item Adaptive grids require us to project the
   solution along resolution transitions onto the finest grid, then to
   solve the formulation there, and finally to restrict the Riemann solve's
   outcome again \cite{Berger:89:AMR}.
   AMR injects dependencies into the Riemann solve phase.
   \item For high flop/s rates, it is important that no
   phase of the solve exhibits low concurrency, has high bandwidth demands or
   synchronises the other tasks. Mesh cells that dynamically refine run risk to do so: If they are
   processed late throughout the sweep, they allocate memory, initialise data
   structures, and then invoke the actual computations, while the other threads might already have run out of tasks.
 \end{enumerate}

\section{Enclave tasking}
\label{section:enclave-tasking}

Our solution to the aforementioned challenges is a technique we label
as {\em enclave tasking.}
It relies on our topological assumption on the DG grid plus one
assumption on typical refinement patterns:

\begin{assumption}
\label{assumption:mesh-changes}
We assume that mesh refinement criteria \added{typically}
refine and coarsen the mesh along resolution transitions:
A cell belonging to grid level $\ell $ might be refined if at least
one adjacent cell has a level $\hat \ell  > \ell $, i.e.~is finer.
A cell belonging to a grid level $\ell $ might be coarsened if at
least one adjacent cell has a level $\hat \ell < \ell $, i.e.~is coarser.
Cells surrounded by cells of the same grid level \added{are typically} neither
\added{refined} nor \added{coarsened}.
\end{assumption}

\noindent
Explicit time stepping schemes for hyperbolic equation systems render our
assumptions on refinement and coarsening  reasonable as the CFL condition
ensures that information does not propagate more than 
one cell at a time.
The assumption \added{does not hold globally} for strongly nonlinear equations
\added{where} areas of interest for a refinement criterion can ``pop up'' as
shocks develop out of smooth solutions.
\added{
 It furthermore breaks down for setups with time-dependent boundary conditions
 or source terms that stimulate a wave throughout the simulation.
 Finally, it does not anticipate that wave spreading can yield large regularly
 refined regions which eventually should make the mesh thin out, i.e.~coarsen
 over a whole subdomain.
}

Yet, it seems that this happens rarely \added{or locally}.
\added{
 In the following, we do not make any semantic modifications to the cell
 treatment.
 We only optimise using the assumption.
 Whenever and where it does not hold, our code does not benefit from the
 optimisations as proposed.
}

 \begin{figure}
  \begin{center}
   \includegraphics[width=0.28\textwidth]{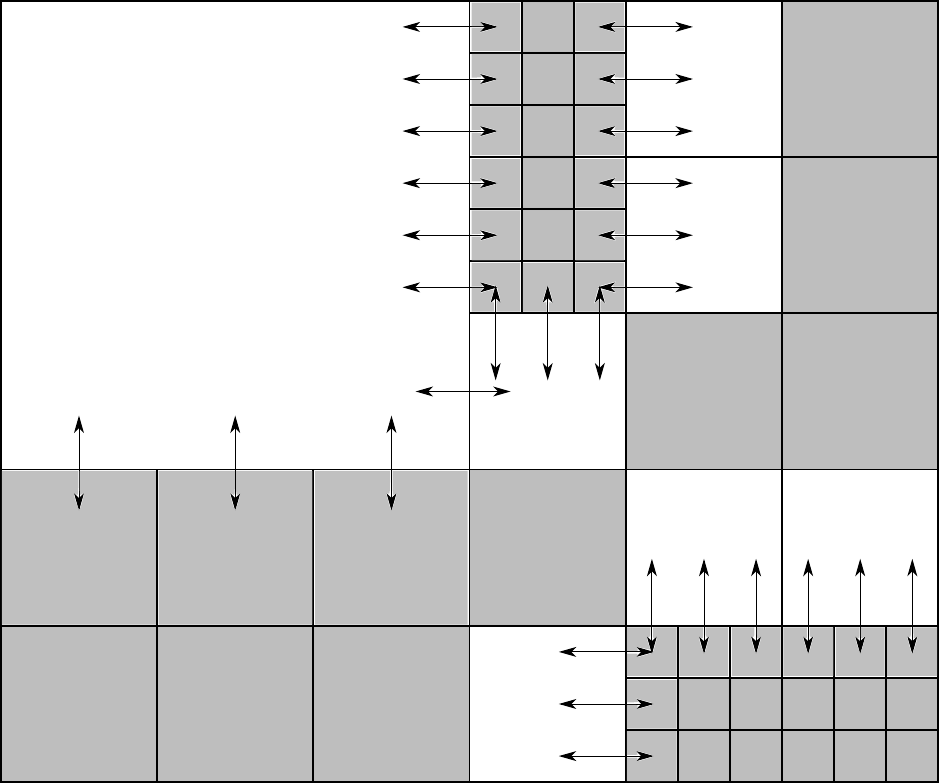}
   \hspace{0.8cm}
   \includegraphics[width=0.4\textwidth]{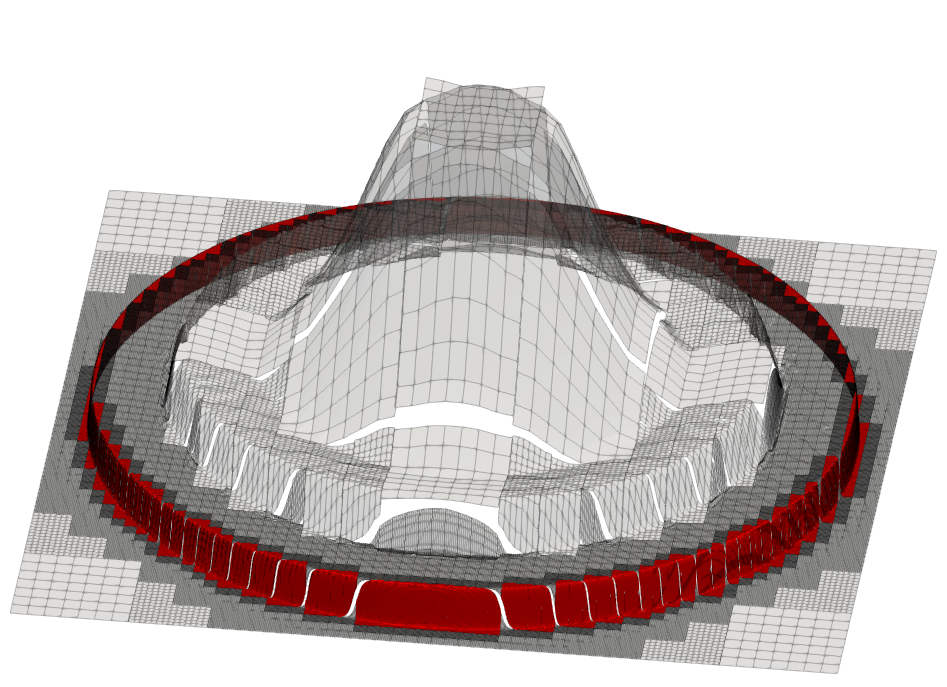}
  \end{center}
   \vspace{-0.2cm}
   \caption{
     Left: An adaptive Cartesian mesh where the Riemann solves along adaptivity
     boundaries are denoted by arrows.
     Cells adjacent to cells of a finer resolution describe a
     skeleton while the filled cells form enclaves.
     Right: Simulation snapshot of an Euler simulation.
     The code uses patch-based FV along the shock (coloured areas)
     and ADER-DG with order $p=7$ everywhere else.
     The gaps in the visualisation are a direct result of the discontinuous
     shape functions.
     \label{figure:enclaves}
   }
%    \vspace{-0.6cm}
 \end{figure}

Let a {\em skeleton grid} of a given adaptive mesh comprise those mesh cells
that are either adjacent to a domain decomposition boundary or are adjacent to
at least one cell of a finer level.
The remaining cells form {\em cell enclaves} (Fig.~\ref{figure:enclaves}).

\subsection{Algorithmic blueprint}

Enclave tasking maps each ADER-DG time step onto two types of mesh traversals.
We refer to them as {\em primary mesh traversal} and {\em secondary mesh
traversal}.
They take turns.
Furthermore, enclave tasking assigns
each cell in the mesh a boolean marker
$STP_{complete} \in \{ \bot, \top \}$.
At construction $STP_{complete}(c)=\top \ \forall c \in \Omega _h$.

\paragraph{Primary mesh traversal}
The primary mesh traversal runs through the mesh.
It satisfies the following properties:
\begin{enumerate}
  \item[P.1] All primary mesh traversals on the parallel computer are
  deterministically reading the faces along MPI boundaries. 
  \item[P.2] A primary mesh traversal reads all $2d$ adjacent faces to any cell
  before it reads the cell itself.
  \item[P.3] A virtual cell is read before the spatially overlapping real
  cell is read.
\end{enumerate}

\noindent
The primary mesh traversal for ADER-DG triggers the following
steps:

\begin{enumerate}
  \item Whenever the traversal reads a face for the first time that 
    is adjacent to an MPI boundary, we receive Riemann solver input data from
    the neighbouring rank. 
    As a result, all data feeding into a Riemann solve is available locally.
  \item For each face read, we check that $STP_{complete} = \top $ for all
    adjacent real cells of the same resolution that are held on the same rank.
    If one flag is not set, i.e.~equals $\bot$, the traversal is suspended.
    We yield.
    The runtime thus gets the opportunity to process other tasks. 
    Upon return, we re-check the condition.
    \label{enumeration:primary-traversal:wait-for-stp}
  \item For each face that is not subdivided further, the traversal computes $F_\text{R}$.
  \item For each child face, the Riemann result is immediately restricted.
    \label{enumeration:primary-traversal:restriction}
  \item For each cell $c$ that we read,
  \begin{enumerate}
    \item we run the corrector,
    \item we evaluate the dynamic adaptivity plus limiter criteria
    \cite{Dumbser:14:Posteriori},
    \item we reset the completion flag $STP_{complete}(c) \gets \bot $, and 
    \item we spawn a new STP task $C_\text{STP}$ if the subsequent time step
    size is known already \cite{Charrier:18:AderDG}.
    \label{enumeration:primary-traversal:stp-spawn}
  \end{enumerate}
\end{enumerate}

The traversal studies the cell's adjacent faces upon its load. If the
cell is adjacent to a resolution transition or adjacent to the MPI domain
boundary, the cell is a skeleton cell.
Otherwise, it is an enclave cell. 
For enclaves, the spawned $C_\text{STP}$ task goes to a standard
task queue.
We prioritise this queue lower than the actual tree traversal and label it
background queue.
Contrary, the task goes into a high priority queue for skeletons.
%\label{enumeration:primary-traversal:wait-for-stp}
% For each enclave cell $c \in \Omega _h$ which carries ADER-DG polynomials

The primary mesh traversal is a task producer that supplies the task runtime with
ready tasks.
The traversal itself can run in parallel.
Semaphores on the faces---which we did not discuss explicitly---ensure that no
race conditions arise from Step \ref{enumeration:primary-traversal:restriction}.
If we know all admissible time step sizes and hence can implement
Step (\ref{enumeration:primary-traversal:stp-spawn}), enclave tasking
logically shifts compute steps, i.e.~brings some tasks forward
\cite{Charrier:18:AderDG}:
The primary traversal runs the Riemann solve plus the two subsequent volumetric
tasks.
The STP among them logically belongs into the next time step.
If this is not possible, we have to run through the grid once more after each
primary sweep and issue the follow-up STPs.
This is an ``unproblematic'' activity from a performance point of view, as STPs
are arithmetically intense.
% The subsequent secondary traversal does not wrap up the time step but solely
% ensures that some $C_\text{STP}$ tasks have terminated.
% It is not until the subsequent primary traversal that the Riemann solves and
% correctors are completed.
% Ammortised, two grid traversals realise one time step.
% Yet, one of them runs solely through the skeleton.
All $C_\text{STP}$ tasks are straightforward realisation of $C_\text{STP}$ from
(\ref{equation:ader-dg:tasks}).
Upon a task's termination, it sets its marker $STP_{complete}(c) \gets \top $.

\paragraph{Secondary mesh traversal}
The secondary mesh traversal runs through the mesh.
It satisfies the following properties:
\begin{enumerate}
  \item[S.1] All secondary mesh traversals on the parallel computer are
    deterministically reading the faces along MPI boundaries.
    If a face separates domain $\Omega _1$ from $\Omega _2$, both ranks $r_1$ and
    $r_2$ owning the respective domains hold a copy. 
  \item[S.2] A secondary mesh traversal reads a cell before it reads/studies any
    of its $2d$ adjacent faces.
  \item[S.3] A virtual cell is loaded after the spatially overlapping real
    cell has been read.
\end{enumerate}

\noindent
The secondary mesh traversal for ADER-DG triggers the following steps:

\begin{enumerate}
  \item For each cell read that belongs to the skeleton, we check whether
    $STP_{complete}=\top$. If not, we yield before we check again.
    \label{enumeration:primary-traversal:mpi}
  \item For each virtual cell that is loaded, we interpolate from its parent and
    we project the interpolated data onto the virtual cell's faces.
    \label{enumeration:primary-traversal:interpolate}
  \item If a face coincides with the MPI domain boundary, we send the $Q^*_h$
    projection from the local adjacent cell to the neighbouring rank.
\end{enumerate}

\noindent
The secondary grid sweep is a degenerated grid traversal traversing
only the skeleton.

\subsection{Relation to trees and forests as well as space-filling curves}

Tree discretisations and traversals fit seamlessly to enclave tasking and its
traversal.
Any coarse to fine traversal \cite{Weinzierl:19:Peano} allows us to
realise it.
%  if the depth-first ordering is used for the primary tree traversal
% and its inverse (backtracking) for the secondary.
If the traversal is realised through a (depth-first) push-back automaton, i.e.~a
recursive function, we embed all routines from the secondary traversal into the
recursive function's preamble before we recurse further (pre-order), while we
realise the primary traversal's steps in the post-order, i.e.~when we backtrack.
For breadth-first, enclave tasking's primary traversal runs
through the grids from fine to coarse.
The secondary
traversal starts with the coarsest resolution.

In a tree world, it is convenient to make the spacetree accommodate both real
cells and virtual cells.
For this, the nodes of the tree are classified as follows:
Inner nodes are refined tree nodes which do not carry an ADER-DG discretisation
but parent further inner or fine grid nodes.
An unrefined fine grid node carries a polynomial from $Q_h$, but is not refined
further.
This is a node that spans a cell of the ADER-DG mesh.
A refined find grid node carries a polynomial, too, but is refined. 
It parents virtual or supplemental nodes but hosts a cell of the ADER-DG mesh,
too, even though it is refined.
A virtual node is unrefined and supports our AMR implementation.
It does not carry a real solution but temporarily is subject to $\hat Q_h^*$
writes.
Supplemental nodes can be refined---along their descendant then are solely
supplemental or virtual nodes---or unrefined.
Their purpose is to complete the tree language
(Fig.~\ref{figure:enclave:tree}).

% The dynamic adaptivity from Step \ref{enumeration:primary-traversal:stp-spawn}
% does not fit directly into a backtracking concept:
% AMR creates finer resolutions which have, by construction, already been
% processed by the primary sweep.
% In this case, it makes sense either to postpone refinement, i.e.~to bookmark a
% refinement request and to realise it in the subsequent secondary sweep, or to
% move the AMR evaluations from the secondary sweep into the primary one.
% We advocate for the latter if a recursive tree traversal is used, as all
% ``plug-in'' points do exist already---it is all the same type of recursion.
% Primary and secondary approach differ only by the fact that functionality is
% either realised after or before the recursion.

In our non-overlapping domain decomposition, ranks compute the Riemann solutions
redundantly.
If a face separates domain $\Omega _1$ from $\Omega _2$, both rank $r_1$ and
$r_2$ owning the respective domains have a copy of this face.
Our scheme assumes that the secondary traversal sends out data. 
These data thus are become available in the subsequent primary sweep on the
destination rank.
To avoid resorting boundary data, it is convenient to make both $r_1$ and $r_2$
traverse their shared faces in the same order or to inverse the order after each
sweep \cite{Weinzierl:19:Peano}.
In these cases, queues or stacks can be used for all boundary data exchange.

\subsection{Properties}

With enclave tasking, the individual ADER-DG steps are not synchronised among
different cells:
Some might still ``wait'' for their Riemann solves and correctors 
\[
  \left( id|_{T+\Delta T} \circ
C_\text{STP} \cdot Q_h(T), id |_{\partial c} \circ C_\text{STP}\, Q_h(T) \right)^T,
\]
while others have already issued $C_\text{STP}\, Q_h(T+\Delta T)$.
The producer-consumer pattern of our traversal ejects ready tasks which can
be ran immediately once cores become available.
The task markers resolve task dependencies.
This is the reason we can work completely task graph assembly free.

%
% MPI
%
A high prioritisation of the skeleton STPs implies that tasks
that yield MPI messages are
processed early.
The secondary mesh traversal then can issue MPI sends, while many remaining
enclave STPs still linger in the ready queue.
We thus give MPI the opportunity to overlap computation and communication and
reduce the risk of a late sender pattern \cite{Mao:14:WaitStateProfiler}.

Expensive inter-resolution transfer operators (restriction and prolongation) are 
either explicitly hidden behind enclave STPs, too, or they are intermixed into the
primary mesh traversal where we may assume that many STP spawns prelude
the first interpolations.
We thus hide their memory-intense operations
behind computations.
Also bandwidth-demanding Riemann solves and correctors mix with computationally
heavy STPs.

%
% Skeletons on the fly
%
Neither the MPI-oblivious behaviour nor the orchestration of task
with different characteristics are constrained by dynamic AMR.
Grids can change in each and every primary sweep.
All skeleton markers are computed on-the-fly.
As mesh refinement is only one substep of the primary traversal, and as the
primary traversal both produces new STP tasks and does not wait for all
STP tasks from the previous time step to complete  before it kicks off,
it is fair to assume that expensive memory allocations, which furthermore
typically struggle to run in parallel, hide behind computations of further
STPs.

%
% Requirements that these characteristics hold
%
The advantageous AMR-agnostic characteristics
of enclave tasking require two assumptions to hold:
On the one hand, the runtime has to get the prioritisation right. 
If the secondary mesh traversal waits for STPs too long and too many
non-critical (enclave) tasks are processed instead, then we will run into
close-to-serial phases in the subsequent primary sweep.
On the other hand, the individual STPs have to be expensive relative
to other algorithmic steps as well as the task production.

\section{Tailoring the task runtime system}
\label{section:tasking}

%
% Rationale behind this realisation
%
Modern task systems are designed to handle millions of small
tasks with dependencies.
For enclave tasking, the latter feature is not required.
Contrary, constructing a dependency graph on-the-fly---the grid
might change every time step---would induce algorithmic overhead/latency that
postpones the processing of the first task.
The spawned STP tasks are ready by construction.
If the grid traversal is deterministic and, besides the AMR grid alterations,
always the same, first-in, first-out (FIFO) task processing 
delivers an optimal task execution order
as long as all skeleton tasks are ran prior to the enclaves.

%\vspace{-0.2cm}
\begin{algorithm}[htb]
  \caption{
    Blueprint of the consumer task.
    \added{It accepts the queue $q$ filled with STPs. These are logical
    tasks not queued into the actual runtime, while the consumer task itself is
    a real task in a multithreading/-tasking sense.
    } 
  }
  \label{algorithm:consumer}
  \begin{algorithmic}[1]
    \Function{runConsumerTask}{task queue $q$}
      \State $C \gets $ \Call{fetch and decrement}{\#consumer tasks}
      \If{$C < size(q) / N_\text{min}$}
       \State \#consumer tasks $\gets $ \Call{fetch and increment}{\#consumer tasks}
       \State \Call{spawn new consumer task}{$q$}
       \State $reenqueue \gets \top$
      \ElsIf{$(C \leq 1) \  \vee \ $ not $empty(q)$ }
       \State $reenqueue \gets \top$
      \Else
       \State $reenqueue \gets \bot$
      \EndIf  
      \State \Call{process up to $N_\text{max}$ tasks}{$q$}
      \If{$reenqueue$}
       \State \#consumer tasks $\gets $ \Call{fetch and increment}{\#consumer tasks}
       \State \Call{spawn new consumer task}{$q$}
       \Comment \added{requeue/re-spawn of the present task}
%        \State \Call{reenqueue consumer task}{$q$}
      \Else
       \State \Call{terminateTask}{}
       \Comment starve one consumer
      \EndIf  
    \EndFunction
  \end{algorithmic}
\end{algorithm}
%\vspace{-0.2cm}

Our realisation wraps around Intel's
Threading Building Blocks (TBB) \cite{Reinders:07:TBB}.
We modify this tasking runtime to accommodate our needs.
Three variants are available.
Our basic variant maps both enclave and skeleton tasks onto native TBB tasks.
The second variant puts the logical tasks into a queue.
It then spawns a number of \emph{consumer tasks} which dequeue these (logical)
tasks and process them.
In our realisation, consumer tasks are the real tasks in the TBB sense. Enclave and 
skeleton tasks are the work items processed by these consumers
(Algorithm~\ref{algorithm:consumer}).
A third variant switches from a plain FIFO queue to a priority queue as provided
by the TBB library.
The high priority/low priority concept of skeletons vs.~enclaves is realised
through priorities (integers) attached to the enqueued items.

% The variants are assessed with some simple Euler equations
% simulating a circular explosion as detailed in the results section.
% We start from $d=2$ meshes on a regular $729 \times 729$ base grid 
% ($\Delta l=0$).
% To this grid, we add $\Delta l \in \{1,2,3,4\}$ additional levels of AMR. 
% The mesh contains
% $531,441$ ($\Delta \ell = 0$), $595,953$ ($\Delta \ell = 1$), $772,209$ ($\Delta
% \ell = 2$), $1,305,490$ ($\Delta \ell = 3$) or $2,902,670$ ($\Delta \ell
% = 4$) cells, respectively.
% We study the polynomials $p \in \{5,9\}$ and cross-validated all data against
% $d=3$ experiments with $\Delta \ell = 0,1$.
% Meshes here contain
% $19,683$, and $139,855$ cells, respectively. 
% The degrees of freedom used by the fifth and nineth order
% approximation are obtained by multiplying the number of cells with $m\,(p+1)^d$, the size of the
% cell solution array, where  is the polynomial order, $m$ the number of state variables,
% and $d \in \{2,3\}$ the dimension of space.

\subsection{Task prioritisation and orchestration}

The efficiency of the task runtime for enclave's producer-consumer pattern
depends on the balancing of task production and task processing.
If the traversal fails to produce enough tasks to keep other cores busy,
performance decreases.
If the processing of heavy STP tasks constrains the traversal, it
runs risk to decrease the performance, too, as the system might run out of
ready tasks later down the line.

%
% If very heavy codes there, then 
%
Our code thus is parameterised through an $N_\text{min}$.
Unless we wait for a termination flag of a STP to be set, the grid
traversal issues at most one consumer task.
Consumer tasks in turn fork into more consumer tasks if the ready queue is
reasonable big, i.e.~if each consumer will have at least $N_\text{min}$ work items.
If the ready queue starts to empty, the consumers starve.
Besides ensuring that there are always enough cores for the tree traversal, such
an approach also spawns new consumer tasks in a binary tree fashion.
The task creation is done by the master thread only for the first
consumer.
The counterpart parameter $N_\text{max}$ ensures that no consumer grabs too many
tasks in a row before it reevaluates the starvation/forking criteria again.

\subsection{MPI progression and MPI buffer layout}

% \vspace{-0.2cm}
\begin{algorithm}[htb]
  \caption{
    The subroutine used to wait for a predictor task to finish.
  }
  \label{algorithm:mpi-progression}
  \begin{algorithmic}[1]
    \Function{waitForSTP}{x}
     \While{x not set}
      \If{\texttt{MPI\_Probe}(any message)}
       \Comment MPI progression
       \State \Call{\texttt{MPI\_Recv}}{}
       \Comment Only one/few messages at a time
      \EndIf
      \State \Call{process up to $N_\text{max}$ tasks}{task queue $q $}
     \EndWhile 
    \EndFunction
  \end{algorithmic}
\end{algorithm}
% \vspace{-0.2cm}

% Status quo
Overlapping communication and computations is important to ensure scalability. 
MPI provides non-blocking routines to this end.
Our secondary grid sweep can trigger non-blocking sends.
The symmetry of the communication---every send out of a face it matched by a
receive---implies that the communication scheme is conceptionally simple.
MPI implementations nevertheless struggle to make the data transfer, the
\emph{data progression}, run in the background
\cite{Hoefler:08:SacrificeThread,Sergent:2018:OverlapCommunicationComputation},
and instead require the user code to poll the MPI subsystem regularly.
This gives MPI a hook in point to manage the actual data transfer.
There are two solutions to realise this polling: 
Either a dedicated progress thread (PT) is deployed, 
or the user code itself calls MPI routines himself.
The latter has to compromise between frequent calls and call overhead,
while most applications do not want to sacrifice a whole thread for MPI
progression only.
With enclave tasking, we can plug into the $STP_{complete}=\top$ checks to progress MPI.
Our main task acts as progression
tasks when it runs into a semaphore (Algorithm~\ref{algorithm:mpi-progression}).
Besides the MPI progression, it also processes some tasks, i.e.~helps out 
on the consumer side.
% The approach is well-suited for non-blocking data exchange.
It inherently overlaps data exchange and computations.

Many MPI codes aggregate MPI data in dedicated buffers before
they send them out.
Each send induces some overhead. 
Message aggregation reduces this overhead.
With enclaves, we however benefit from small messages:
% Skeleton STPs send out Riemann input data in the secondary sweep as small
% chunks.
Enclave partitioning ensures that partition domain boundary data are
sent out early compared to work done in the interior.
Throughout waits for STP results, we receive Riemann input data chunk by
chunk. 
Exchanging small chunks of non-aggregated per-face data ensures that no single receive
of a very large message delays the simulation progress.
\added{
 As we ensure that all MPI data are received in the right order
 \cite{Bungartz:06:Parallel,Weinzierl:19:Peano}, and thus avoid both resorting
 overhead and too many unexpected messages.
}

\section{Generalisation of enclave tasking}
\label{section:dg}

ADER-DG is a peculiar explicit time stepping scheme. 
We use it as showcase for enclave tasking.
However, the enclave concept applies to a variety of DG approaches for a
variety of problems.

\subsection{Explicit Runge-Kutta DG schemes}
\label{section:dg:Runge-Kutta}

For traditional explicit DG schemes including Runge-Kutta, the weak
formulation of (\ref{equation:introduction:PDE}) yields operations
where neither the volumetric integrals (tasks) feed into the Riemann solves nor the
other way round.
We however have to bring together the Riemann and volume integrals  outcomes 
to construct a subsequent time step
$Q_h(\cdot,T+\Delta T) = \left( F_\text{R} + C_{DG} \right) Q_h(\cdot,T) $ or an intermediate step in
the Runge-Kutta tableau.
We therefore propose to make the Riemann solves feed logically into the
volumetric kernels:
A volumetric kernel computes the weak formulation over the cell, but it also
accepts the outcome of the $2d$ adjacent Riemann solves and immediately merges
them into the result.

In such a setup, enclave tasking requires two grid sweeps:
A primary traversal computes all Riemann problems.
It also issues all cell tasks bringing the ingredients together.
As a cell is read after its $2d$ adjacent faces have 
been read, all cell tasks are by definition ready.
The secondary traversal degenerates.
It does not compute anything anymore, but projects all updated solutions onto
the faces immediately such that they are sent out to adjacent ranks and
available there in the next primary sweep.

\subsection{Finite Volumes}

Finite Volumes for explicit time stepping schemes choose piece-wise
constant shape functions for $Q_h$.
They thus can be read as ADER-DG scheme with a degenerated STP.
Solely the Riemann outcome determines a subsequent time
step $Q_h(\cdot,T+\Delta T) = \left( F_\text{R} + id \right) Q_h(\cdot,T) $.
Enclave tasking relies on reasonably expensive STPs
such that grid traversal and Riemann solves disappear behind all the volumetric
tasks.
Straightforward FVs are a bad fit to enclave tasking.
%  in a straightforward implementation.

Yet, all depends on the realisation of the Riemann solve.
Many advanced FV solvers rely on a sophisticated reconstruction of the solution
that they feed into the Riemann solver.
If such a construction decomposes into a ``left'' and ``right'' contribution
that we can compute independently of each other, these reconstruction steps can be
outsourced as volumetric kernels:
Each cell task computes the respective reconstructions for $2d$ Riemann solves
on its adjacent faces.
If such outsourcing is possible, enclave tasking can help:
A primary sweep triggers all reconstructions, a secondary sweep ensures that
reconstructed data is sent over the network and AMR is handled properly, and the
subsequent  final primary sweep---which we might combine with the next time step
\cite{Charrier:18:AderDG}---then issues the actual Riemann solve.

\subsection{Block-structured methods}

It is this discussion of volumetric cost vs.~face tasks that implies that
block-structured AMR \cite{Dubey:16:SAMR} and enclave tasking fit together. 
In block-structured AMR, blocks or patches---typically regular Cartesian
grids---are embedded into the cells.
They communicate with their neighbours through halo layers.
In such a scheme, the Riemann tasks $F_\text{R}$ becomes a halo layer exchange task and
we end up with the situation described before where the face tasks feed into the
volumetric updates.
As long as the halo updates are cheap compared to the patch updates, enclave
tasking is of value.

We use block-structured FV in our own ADER-DG code as limiter
\cite{Dumbser:14:Posteriori}:
Our code determines the solution update through ADER-DG. 
If the resulting solution is physically wrong---if they yield negative
densities, e.g.---or if the solution exhibits oscillations, we roll back ADER-DG
on the respective cell and replace the time step for this particular cell with a
FV scheme.
To match the ADER-DG time step for a polynomial ansatz $p$ with the FV time
steps, the patches per cell have dimension $(2\,p+1)^d$.
Though this solver hybrid validates the claim that (block-structured) FV
benefits from enclave tasks, another pro-enclave argument has to be read with
care:
Enclave tasking for DG ensures that the bandwidth- and latency-sensitive face
tasks dribble through the system and that the runtime orchestrates compute-heavy
volumetric tasks around them.
In an FV world, patch updates tend to be bandwidth-bound. 
The orchestration argument collapses for a pure FV approach.
It continues to hold for ADER-DG where the FV is an a posteriori limiter. 
Here, we may assume that only few cells from the domain are limited,
i.e.~classic ADER-DG cells remain and their volumetric kernels now mix with
Riemann solves, traversal tasks and the patch-based FV updates. 

\subsection{Implicit schemes and linear equation system solves}

Iterative linear equation system solves for DG as they arise for elliptic
problems and implicit time stepping schemes typically rely on matrix-vector
products over (\ref{equation:introduction:PDE}).
They thus resemble the situation of Runge-Kutta schemes from Section
\ref{section:dg:Runge-Kutta}.
Enclave tasking thus can be of value if we work matrix-free
\cite{Weinzierl:18:BoxMG}.
Particular appealing is the combination with multiplicative $hp$-multigrid.
Here, the fine grid smoothing and, hence, residual computation is the dominant
step.
If we run multiple smoothing steps in a row which are followed by
a final residual computation that feeds into a restriction, enclave tasking
unfolds its full potential to hide the Riemann solves behind the volumetric
kernels.

For many elliptic (sub-)problems, codes
start with initial meshes that resolve sources of interesting
behaviour---typically material transitions or complex boundaries---accurately
right from the start.
The areas of interest are known, and the code develops the AMR mesh from there.
Errors from ``problematic'' regions
decay from there according to their fundamental solution, i.e.~the finer grids
of an appropriate mesh follow this decay.
They spread from the problematic region.
Refinement criteria can be throttled to refine at most one additional
layer around a given region of interest per iterate.
This makes a dynamic
refinement pattern fit to our AMR assumption.

\subsection{2:1 balancing and $k$-partitioning}

\added{
 Many codes or numerical implementations require 2:1 balancing
 \cite{Sundar:08:BalancedOctrees}, while many refinement criteria yield
 reasonably balanced grids automatically.
 Our definition of enclaves and skeletons does not rely on a 2:1
 balancing property.
 It is agnostic of resolution balancing.
}

\added{
 If grids are balanced plus feature large resolution transitions, we observe a
 higher skeleton-to-enclave cell ratio than for a mesh with the same
 resolution difference yet no balancing at all.
 In balanced meshes, refined regions spread out gradually---they
 ``ripple'' through the domain---where other codes would feature massive
 resolution jumps over $\Delta \ell$ grid levels.
 Where the latter exhibit one fine--coarse transition manifold, a balanced grid
 features $\Delta \ell$ of these transitions hosting skeleton cells.
 Bipartitioning amplifies this skeleton impact.
 With bipartitioning, whole transition regions can become skeletons.
 For grids featuring $k\geq 3$-subdivision
 \cite{Weinzierl:11:Peano,Weinzierl:19:Peano}, transition skeletons are always interrupted by enclaves.
}

\added{
 We hence may assume that bipartioning and balancing diminish the performance
 gain through enclave tasking. 
 Unbalanced grids with $k \geq 3$-subdivision benefit more from it.
 The analysis and validation of this hypothesis is however out of scope
 here.
}

\section{Experimental results}
\label{section:results}

We benchmark our algorithm and our code on Super\-MUC-NG
at the Leibniz Supercomputing Centre
(LRZ).
Its nodes are two-socket systems, i.e.~each node hosts two 24-core Intel Xeon 8174 (Skylake) CPUs.
They have been clocked at 2.3~GHz and are connected through Intel Omni-Path.  
We have 96~GB main memory available on each node. 
% The vendor specifies a maximum memory bandwidth of 102.4 Gbyte/s,
% at a level 2 cache size of 256 kByte per core,
% a level 2 cache bandwidth of 42 GByte/s per core,
% a level 3 cache size of 2x20 Mbyte, 
% and a level 3 cache bandwidth of 31 GByte/s per core.

Enclave tasking is a generic concept within the ADER-DG mindset
but applies to FV as special case of DG schemes, too.
We integrated our ideas into the ExaHyPE \cite{Software:ExaHyPE,Reinarz:19:Exahype} 
engine where they support ADER-DG for orders $p \in [3,\ldots,9]$ but also
a patch-based Finite Volume scheme which is used by
ExaHyPE\footnote{\url{www.exahype.org} \cite{Software:ExaHyPE}.} for
a-posteriori limiting \cite{Dumbser:14:Posteriori}.
All results are thus obtained with applications built upon ExaHyPE.

We study the enclave impact for two applications with 
different character.
One application is a seismic wave code that solves the
LOH.1 benchmark.
The cuboid domain used in this benchmark consists of two 
material layers. Wave propagation is initiated by a point source
that is placed just below the upper layer.
Due to the material transition, an interesting wave pattern emerges
(Fig.~\ref{figure:enclave:tree}).

This well-known benchmark is governed by a linear variant of
(\ref{equation:introduction:PDE}).
However, we translate it into a nonlinear variant where the material enters the
equation as an additional scalar PDE over $\alpha (t)$ following the trivial
rule $\partial _t \alpha = 0$.
It does not move.
Such an immersed boundary approach allows us to handle a material transition which
is not grid aligned \cite{Tavelli:19:ImmersedBoundary}:
Wherever the code encounters a transition, we cover it with an FV patch.
Some distance away from the boundary, these FV patches are 
coupled with ADER-DG cells.
In the majority of the domain, we thus use ADER-DG.
Though phrased overall as nonlinear PDE---the $\alpha $ term injects this
nonlinearity---the code's high-order ADER-DG degenerates to a
linear case within the majority of the domain.
The STP thus is directly solved through Cauchy-Kowalevski.
Only on the FV patches along the material transition, we 
solve the original nonlinear PDE.
The material plus the nonlinear PDE require us to store 13 doubles per degree of
freedom.

Our second benchmarks is prescribed by the   
compressible Euler equations \cite{LeVeque:02:FiniteVolumes}.
Compared to the seismic setups, this setup has only five unknowns (a scalar
material density and a scalar energy which closes the system plus the vector of velocities).
However, it is a nonlinear variant of (\ref{equation:introduction:PDE}),
which does generally not degenerate to a linear case and thus 
requires an expensive nonlinear STP where we do not know the number of internal Picard iterations a priori.
This solver switches to FV as a posteriori limiter if
shocks are encountered:
the solver uses ADER-DG with high order in the majority of the domain but
employs a patch-based FV scheme along discontinuities
\cite{Dumbser:14:Posteriori}.
Different to our immersed boundary setup, the FV regions travel this time.
We always refine the DG solution down to the FV level close to the shocks and
then glue ADER-DG and FV together volumetrically.
This implies that no adaptivity cuts through the FV subdomains, and FV 
cells by definition thus are enclave cells.

We ran all experiments in 3d.
\added{
 The timings are given per time step, i.e.~we freeze particular adaptivity
 patterns throughout the measurements.
 Efficient variants of tree modifications including fast balancing---if
 required---are known
 \cite{Isaac:15:RecursiveAlgorithmsOnOctrees,Sundar:08:BalancedOctrees} and
 suggest that the total time-to-solution character of a simulation does not change dramatically.
 Alternatively, any remeshing as well as propagation of the limiter within the
 mesh can be integrated into the actual time stepping along the lines of 
 \cite{Charrier:18:AderDG,Weinzierl:18:BoxMG}.
 With dedicated remeshing phases, enclave tasking does not impact these phases'
 runtime, while Assumption \ref{assumption:mesh-changes} holds in a hard sense,
 i.e.~we can omit the phrases \emph{typically}:
 Throughout the time steps, the skeleton labels never change.
 We can determine the skeleton vs.~enclave classification once throughout 
 the mesh adaption and then keep them until the mesh changes again.
 With a merger of time stepping and AMR, enclave tasking might deteriorate
 locally and we have to update the markers on-the-fly for each and every time
 step.
}

\subsection{Computational characteristics}

\begin{table} 
 \caption{
  \noindent
  \added{Serial runtime}
  in seconds per ADER-DG solution degree of freedom per time step per task type for
  polynomial orders $p \in \{3,5,7\}$. 
  \label{table:cost-per-dof-per-phase}
 }
 \vspace{-0.3cm}
 {\footnotesize
 \begin{center}
  \begin{tabular}{l|rr|rr}
   & \multicolumn{2}{c|}{Seismic} 
   & \multicolumn{2}{|c}{Euler} \\ 
   & \multicolumn{1}{|c}{ADER-DG}
   & \multicolumn{1}{c}{FV}
   & \multicolumn{1}{|c}{ADER-DG}
   & \multicolumn{1}{c}{FV}
   \\
   \hline
   $p=3$, $C_\text{STP}$   
%    & $5.95\cdot10^{-7}$ 
    & $1.22\cdot10^{-6}$ & --                  & $2.72\cdot10^{-7}$ & --        
    \\
   $p=3$, $F_\text{R}$     
%    &   $1.80\cdot10^{-8}$ 
    & $2.42\cdot10^{-8}$ & --                  & $3.59\cdot10^{-9}$ & --        
    \\
   $p=3$, $C_\text{Corr}$  
%    &   $7.25\cdot10^{-8}$ 
   & $6.98\cdot10^{-8}$ & $6.89\cdot10^{-5}$  & $2.74\cdot10^{-7}$ & $2.76\cdot10^{-5}$    \\ 
   \hline                                               
   $p=5$, $C_\text{STP}$   
%   &   $6.01\cdot10^{-7}$ 
   & $1.75\cdot10^{-6}$ & --                  & $3.42\cdot10^{-7}$ & --                    \\
   $p=5$, $F_\text{R}$     
%    &   $1.50\cdot10^{-8}$ 
    & $1.76\cdot10^{-8}$ & --                  & $2.04\cdot10^{-9}$ & --        
    \\
   $p=5$, $C_\text{Corr}$  
%    &   $1.00\cdot10^{-7}$ 
   & $8.06\cdot10^{-8}$ & $6.48\cdot10^{-5}$  & $1.63\cdot10^{-7}$ & $2.70\cdot10^{-5}$    \\ 
   \hline                                               
   $p=7$, $C_\text{STP}$   
%    &   $5.29\cdot10^{-7}$ 
   & $2.79\cdot10^{-6}$ & --                  & $6.07\cdot10^{-7}$ & --                    \\
   $p=7$, $F_\text{R}$     
%    &   $7.95\cdot10^{-9}$ 
   & $1.20\cdot10^{-8}$ & --                  & $1.29\cdot10^{-9}$ & --                    \\
   $p=7$, $C_\text{Corr}$  
%    &   $8.89\cdot10^{-8}$ 
   & $8.80\cdot10^{-8}$ & $6.23\cdot10^{-5}$  & $1.70\cdot10^{-7}$ & $2.65\cdot10^{-5}$    \\
  \end{tabular}
 \end{center}
 }
% \vspace{-0.6cm}
\end{table}

%
% What we do
%
Prior to algorithmic studies, we benchmark how much time we have to invest into
one degree of freedom time step update. 
This cost is broken down into the ADER-DG or FV cost
(Table \ref{table:cost-per-dof-per-phase}).
For our hybrid codes combining ADER-DG and FV, the solver in practice will yield
a mixture of the two characteristics as both codes run concurrently.
We furthermore emphasise that the total runtime cost of a simulation will
comprise grid management and parallelisation overhead as well as adaptive
meshing cost.
The latter comprises the evaluation of refinement and coarsening criteria,
interpolation and restriction.
Refinement criteria evaluation and inter-resolution transfer operators are 
fused with the correction steps in our code. 
The remaining overheads are negligible.

%
% What we see
%
We clearly see that the two ADER-DG solvers have different solver
characteristics:
The ratio between STP and Riemann solve is comparable, but the dynamic
AMR is expensive.
As we merge the latter into the STP in our code, 
Euler's cell tasks are significantly more expensive relative to the 
Riemann solves than the seismic cell tasks.
In contrast, the FV patch updates are by magnitudes more
expensive than all the ADER-DG cells.
For all setups, the ratio of the  STP cost to the remaining
tasks shifts towards the STP with rising $p$.
We conclude the data interpretation with a remark that our ADER-DG scheme is
aggressively optimised towards Intel architectures, whereas the FV scheme is
relatively straightforward. 
For many applications, it might be possible to reduce its cost per degree of
freedom.
This fact is beyond scope here.

%
% Interpretation
%
With these results, we expect the genuine nonlinear PDE (Euler) to benefit more
significantly from enclave tasking than the seismological application
which is effectively linear in the majority of the domain.
In general, we expect the impact of enclave tasking to
become more significant as we increase the polynomial order.
It is clear that enclave tasking should notably become very important once we
have very expensive enclave tasks. 
This is the case for our FV cells.
It is however not clear how their non-predictability (we do not know a prior
where limiting is required) affects the runtime and performance of the scheme.

% In line with our theory, enclave tasking's impact should be the higher the less
% often the FV scheme is employed though the exact mechanisms remain unclear as
% discussed.

\added{
 Our code employs relatively simple Riemann solvers. 
 More expensive solvers shift the emphasis away from the STPs and
 thus diminish the impact of enclave tasking, unless
 the Riemann solve is thinned out:
 For many solver variants, only the reconstruction step bringing data from the
 adjacent cells together has to be realised
 within the face-associated compute kernels, while the actual Riemann solve can
 be outsourced to the corrector.
 Such an implementation breaks the logical face operation up into a face part
 and a volumetric computation.
 The latter can be merged into the corrector (see the discussion on lifting
 \cite{Klockner:09:DGonGPUs} in the FV context).
 It is an open question to which degree the arising increase of compute
 load---in many cases computations of the volumetric part of the Riemann solve
 will run redundantly on both adjacent cells of a face---is compensated by an increase of
 the efficiency of the enclave tasking as well as by the increase of data access
 locality due to the merger of Riemann operations and correction.
}

\subsection{Impact of the task runtime parameters}

A second assessment studies the behaviour of our tasking system, i.e.~it
benchmarks the tasking runtime's tailoring against native TBB.
% For the time being, we omit the tasking's impact on the distributed memory
% parallelisation, as we focus solely on technical, single-node effects.
All tests are run for the two-dimensional Euler equations simulating a circular
explosion.
We employ a $729 \times 729$ base grid.
If we activate dynamic AMR, this setup yields rather aggressive, time-dependent
mesh adoptions.

\begin{figure}
 \begin{center}
 \includegraphics[width=0.45\textwidth]{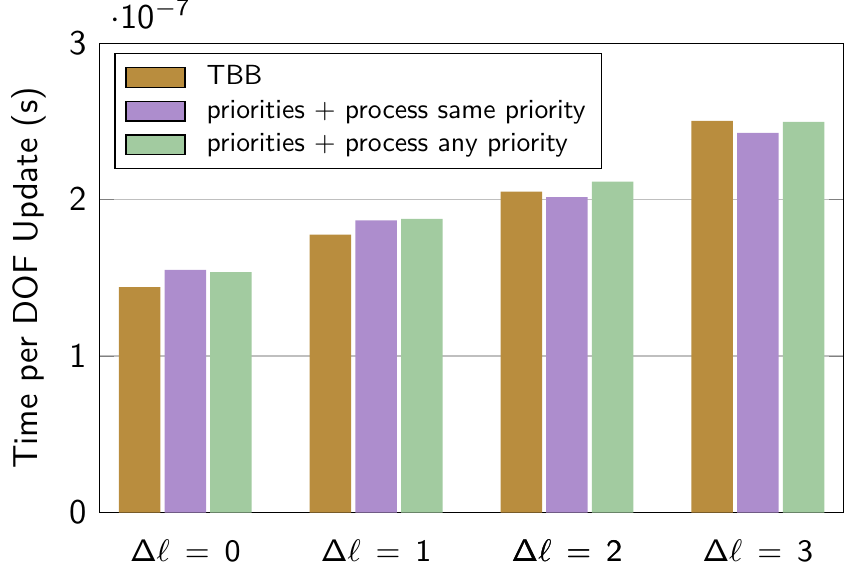}
 \hfill
 \includegraphics[width=0.45\textwidth]{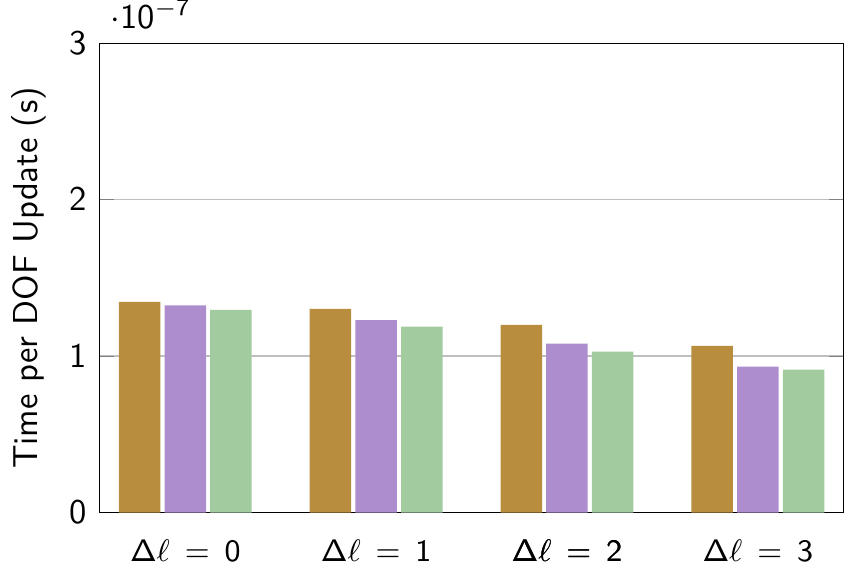}
 \end{center}
 \vspace{-0.4cm}
 \caption{
  Impact of different task processing strategies for $p=5$ (left) and $p=9$.
  All strategies except TBB use $N_\text{min}=N_\text{max}=8$.
  $\Delta \ell=0$ denotes a regular grid, $\Delta \ell$ otherwise denotes the
  maximum number of added AMR mesh levels.
  \added{
   One socket is used.
  }
  \label{fig:tasking:job-processing-strategies}
% \vspace{-0.6cm}
  }
\end{figure}

%
% What we do
%
Our first set up tests fixes the quantities $N_\text{min}=N_\text{max}=8$,
i.e.~whenever a thread processes STP tasks, it tries to process eight tasks in a
row.
The experiments use one socket of the two-socket system to exclude NUMA
phenomena.
They start from a regular grid and add up to three levels of dynamic AMR.
We benchmark native TBB where every STP task is spawned as a real TBB task
against implementations that put these STPs into a priority queue and create TBB
(consumer) tasks which grab them from there.
Skeleton tasks have higher priority than enclave tasks.
% Further to that, we have a priority queue variant where the grid traversal
% processes STPs itself whenever it has to wait for STP outcomes. 

%
% What we see
%
Our data (Fig.~\ref{fig:tasking:job-processing-strategies}) suggest that the
queue wrap-around induces a non-negligible overhead.
Plain TBB is thus faster than any modifications if the computational load per
STP is sufficiently low.
However, aggressive AMR or high relative STP cost imply
that an anarchic spawning of native TBB tasks leads into situations where
either the main thread becomes idle as it waits for an STP to finish, or where
the tasking system unfortunately processes the wrong, i.e.~enclave, tasks, or
where the main traversals tasks yield, their threads pick up other enclave tasks
and eventually return too late to the actual traversal such that we face delays
or work starvation later down the line.
The grid traversal threads should process STPs themselves
whenever they wait for an STP outcome.
If they do so, they naturally prioritise skeleton tasks.
If no more skeleton tasks are ready and the STPs are heavy, it is advantageous
to switch to enclave tasks.
This happens if all skeleton tasks are currently processed.
If the STPs are not that heavy, it is however better to make a thread wait
for STP outcomes, i.e.~to actively poll the completion, whenever no enclave
tasks are remaining.
Again, actively joining the STP computations for enclaves introduces delays down
the line.

Further experiments (not shown) demonstrate that a switch to three dimensions,
other polynomial orders or other applications does not alter our observations
qualitatively:
It is the presence of dynamic AMR and the relative cost of the STPs that
determine which task processing strategy is the fastest.
Yet, switching to three dimensions or the activation of a limiter increases this
relative cost and thus moves the turnover points.
Experiments with various $N_\text{max}$ and $N_\text{min}$ values support the statements
on the overhead. With the best-case processing strategy from above, 
the two magic parameters make a difference for relatively cheap STPs; that is
low polynomial order $p$ for Euler.
Here, it is advantageous to choose reasonably big $N_\text{min} \approx 8$.
A value of eight logically fuses tasks and thus reduces overhead. 
$N_\text{max}$ plays no major role for regular meshes.
Yet if we tackle a rather adaptive mesh, it is better to have an $N_\text{max}$
close to $N_\text{min}$ to allow the task processing to re-evaluate the queue often.
% Overall, the impact of $N_\text{max}$ and $N_\text{min}$ is however not significant; 
For high relative STP cost, $N_\text{max}$ and $N_\text{min}$ seem to play no role.

%
%  Lessons learned
%
Our data suggest that fine-granular prioritisation is an important feature of
(future) task systems and that this prioritisation---different to our manual
approach---should come along with low overhead.
It is obvious that the impact of prioritisation depends on the relative cost of
tasks:
It is this cost that might render non-prioritised, i.e.~less sophisticated,
scheduling superior. 
Future task systems will have to investigate into a balancing of overhead
vs.~optimality.
The other interesting balancing observation is that the best-case scheduling
seems to depend on the grid regularity.
A homogeneous task spawn pattern asks for a different task processing than a
strongly irregular pattern resulting from dynamic AMR.
This effect is also worth studying from a task system's point of view.
As high-order, three-dimensional setups that utilise AMR are of primary
interest to us in this study, we stick to our custom-made tasking
with prioritisation and process-tasks-if-you-wait strategy from hereon.
Low order schemes are typically only used in our setups when we strive for very
aggressive AMR.
We thus set $N_\text{max} = N_\text{min} = 8$.

\subsection{Shared memory scaling}

Enclave tasking can be read as on-the-fly sorting of tasks while they drop
in, where the sort heuristic is guided by the grid adaptivity pattern and the
parallelisation.
It brings time-critical tasks forward.
% The evaluation of this property starts with shared memory runs.

\begin{figure}[htb]
 \begin{center}
  \includegraphics[width=0.325\textwidth]{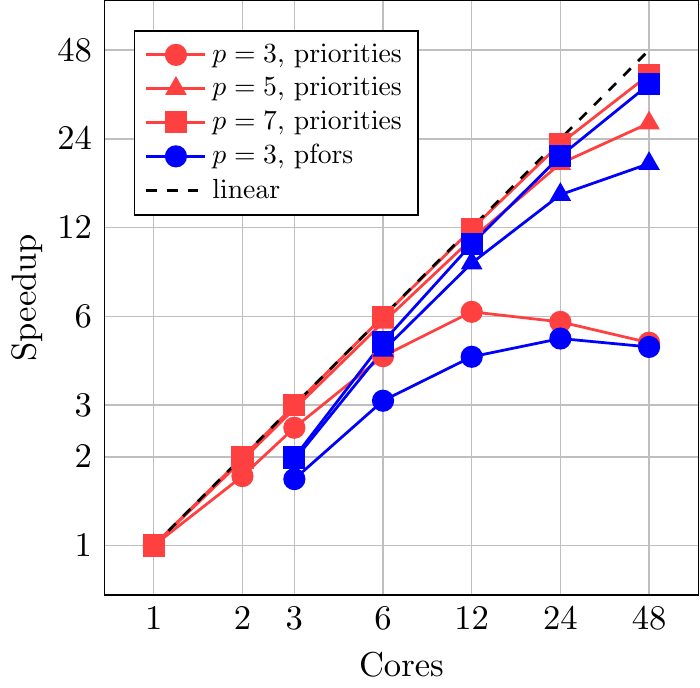}
  \hspace{1.2cm}
  \includegraphics[width=0.325\textwidth]{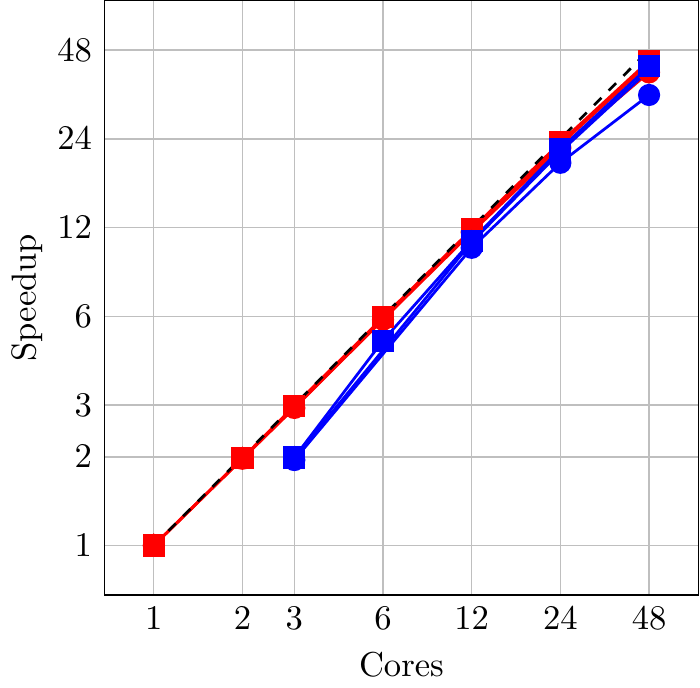} 
  \\ 
  \includegraphics[width=0.325\textwidth]{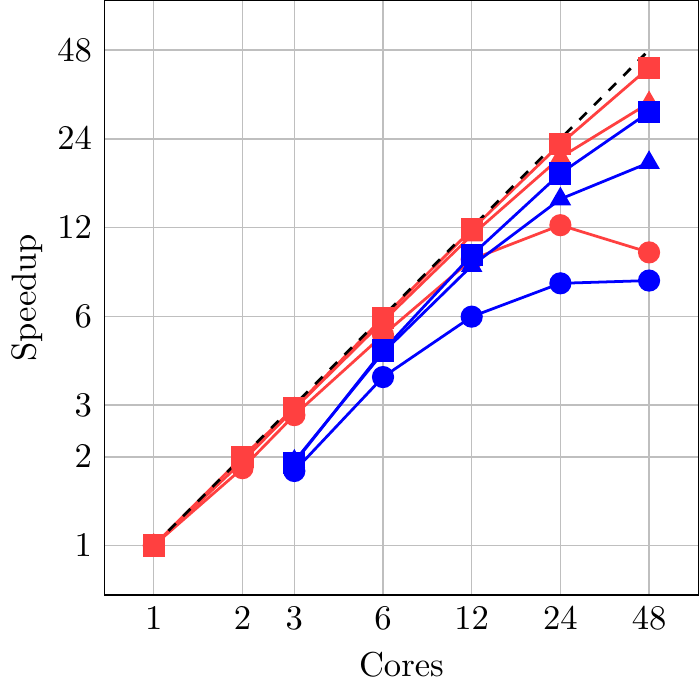} 
  \hspace{1.2cm}
  \includegraphics[width=0.325\textwidth]{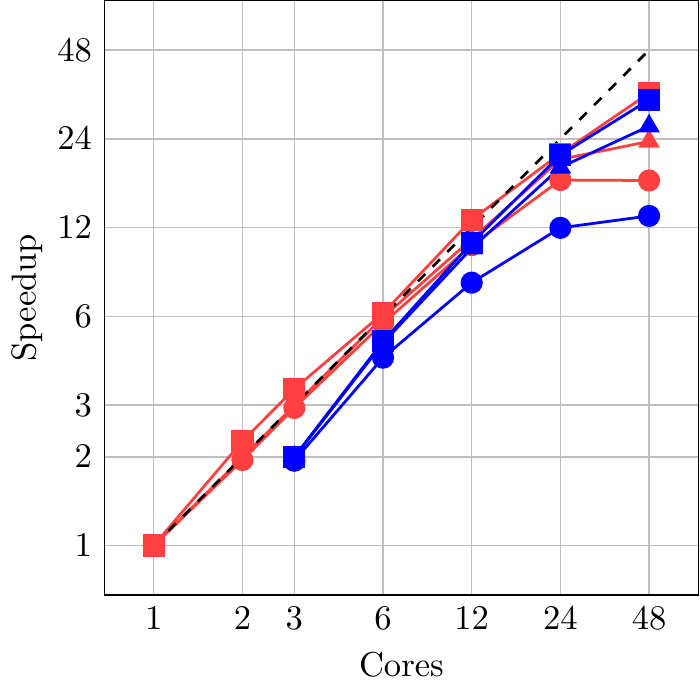}
 \end{center}
 \vspace{-0.4cm}
 \caption{
  Shared memory scaling for regular mesh, i.e.~without any adaptivity
  \added{(strong scaling)}.
  Top row: 
  We compare ADER-DG Euler with smooth initial conditions (left) against
  FV (right).
  Bottom row:
  ADER-DG with a FV limiter simulating Euler with shocks (left) against
  the immersed boundary method of the seismic simulation (right).
  \added{
    All plots study three different polynomial orders $p$ (circle, triangle,
    square) and benchmark classic parallel for-based parallelism against our
    enclave tasking with priorities (red vs.~blue).
  }
  \label{figure:experiments:shared-memory:regular-mesh-runs}
 }
% \vspace{-0.6cm}
\end{figure}

%
% Description: regular mesh.
%
We benchmark our code first on a regular grid in shared memory mode
(Fig.~\ref{figure:experiments:shared-memory:regular-mesh-runs}).
Our baseline is compiled  without TBB.
For Euler, we distinguish between smooth and non-smooth initial
conditions, and we benchmark ADER-DG and limiting ADER-DG
against patch-based FV.
It is only for non-smooth initial conditions (with shocks) that ADER-DG for
Euler uses FV as limiter.
In this case, the scheme becomes a hybrid of both solvers.
With smooth initial conditions, no limiter is required.
It remains sole ADER-DG.
Whenever FV is used, our patch size is chosen as $(2\,p+1)^d$ relative to the
corresponding ADER-DG scheme.
This guarantees that the CFL condition yields time step
sizes of matching magnitude.
Different to Euler, 
the immersed boundary approach uses a limiter always, yet only along
the immersed boundary.
Its limited region stays in place,
while the Euler equations move the limited regions along the shock.
For all tests, a regular grid allows us to 
compare our enclave implementation to a straightforward implementation
with parallel fors.
The latter realises ADER-DG as a sequence of three loops triggering STP,
Riemann solve and corrector.

%
% What we see
%
% (Euler for smooth initial conditions with $p=3$) 
All setups besides the one with the very low relative STP cost
scale reasonably if we use parallel for loops (\texttt{pfors}).
We omit two-core results for TBB's parallel for,
as TBB sacrifices one hyperthread for the scheduling of the loop.
This kick-off penalty plays no major role for higher core counts.  
%, though
%there is some overhead to pay for switching from a serial code to a
%two-core version.
% From two cores on, we see robust upscaling.
% The superlinear results stem from an effective usage of the second memory
% controller.
% Higher core results still
% remain bounded by a linear speedup normalised to a two-controller baseline.
One-to-two overheads do not arise for enclave tasking.

Enclave tasking is robustly faster than loop-based parallelism.
This difference is more significant if the limited region changes over time.
If we use high orders 
%plus limiting 
which implies that the limited region within the grid changes infrequently
relative to the time steps or if we use expensive cell updates (FV or high
orders), both parallel for and enclave tasking play in the same league.

%
% Interpretation
%
% The cache behaviour is one significant showstopper as soon as we enter the
% second socket \cite{Charrier:18:RSC}.
% This property holds for the parallel loop, too, which is not immune against the
% fact that the Riemann solves access the memory non-continuously if we block the
% volumetric data.
% 
On a regular grid, the loop-based parallelism scales excellently for the STPs.
The arithmetic intensity for the two subsequent steps however is not high.
It diminishes the overall scalability.
If we run high orders, the runtime of the STP becomes so dominant that
the impact of these other steps disappear.
For smaller orders, it is significant.
With enclave tasking, this effect however is hidden behind the STPs:
Our approach to orchestrate the tasks such that Riemann solves and STPs
overlap and the Riemann solves dribble through the system with restricted
concurrency pays off.
% ; in particular where the simple parallelisation---though using
% dynamic scheduling---struggles to cope with the fact that the cost per cell
% update is neither predictable as a limiter might require reruns nor statically
% distributed.
% The limited regions travel with the wave.
% Such non-uniform cost setups with significant inhomogeneity are advantageous for enclave tasking as there is no
% hard post-STP synchronisation.

\begin{figure}[htb]
 \begin{center}
  \includegraphics[width=0.325\textwidth]{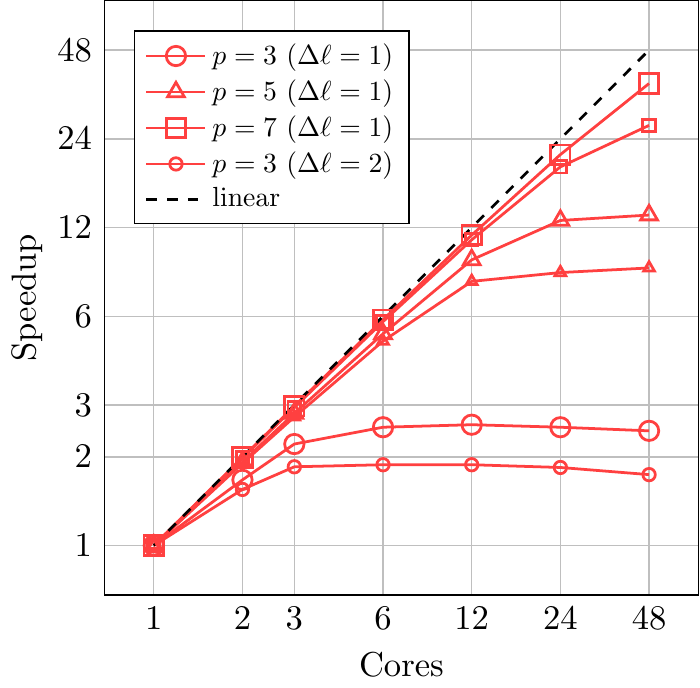} 
  \includegraphics[width=0.325\textwidth]{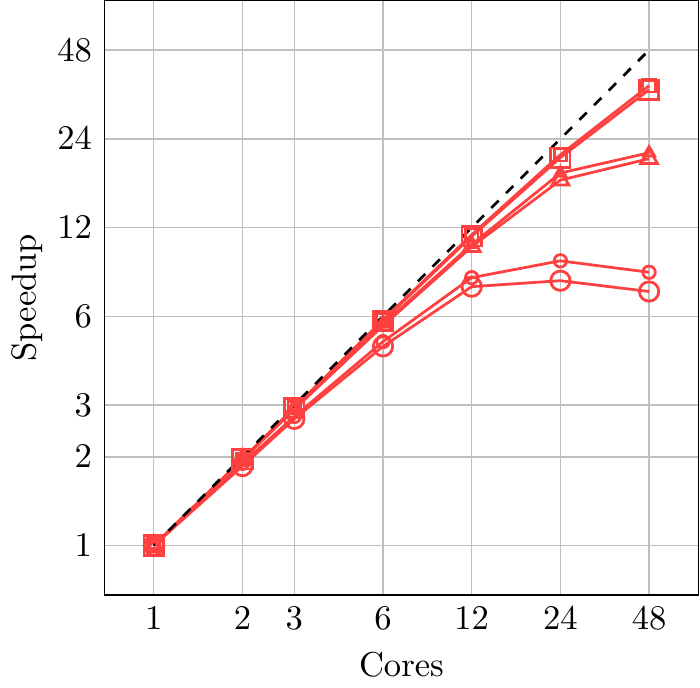}
  \includegraphics[width=0.325\textwidth]{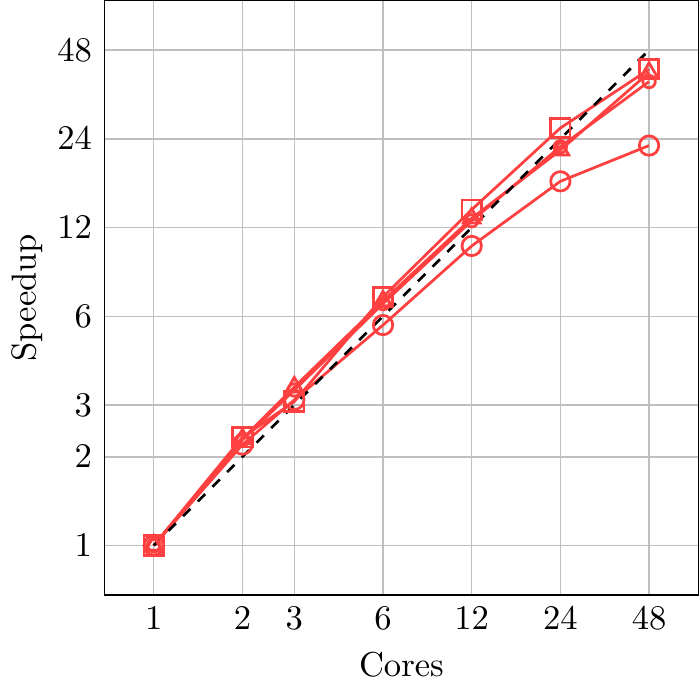} 
 \end{center}
 \vspace{-0.4cm}
 \caption{
  Shared memory scaling with \added{dynamically} adaptive meshes. We present
  data for Euler with smooth initial conditions (left) and Euler with shocks and, hence, limiting
  (middle) as well as immersed boundary data for the seismic setup (right).
  \added{
   Different symbols are used for different polynomial orders $p$.
   The smaller symbol size represents a ``more adaptive'' mesh featuring
   resolution transitions of $\Delta \ell = 2$ levels.
   All three diagrams follow the same symbol semantics.
%    Missing data (right) indicates that the node ran out of memory.
  }
  \label{figure:experiments:shared-memory:adaptive-meshes}
 }
\end{figure}

%
% What we do
%
We continue with adaptive meshes.
The adaptivity for the immersed
boundary method is static, while the adaptive pattern moves along the waves for
Euler. Our adaptivity thus is two-fold: there is adaptivity in space and adaptivity
in the solver. Since a loop-based parallelisation of dynamically adaptive meshes is not
trivial, we omit comparisons to parallel loops.

%
% What we see	
% 
The seismic setup's 
static adaptivity does not pose any problems to enclave tasking. 
In line with the regular grid tests, the $p=3$ tests with only one adaptivity
level are the only ones which fail to yield an arithmetic intensity that leads
to close-to-linear speedups.
Experiments with $p>3$ plus more than one level of adaptivity are impossible due
to memory restrictions, but all remaining
data are more or less AMR-agnostic.
For a sole ADER-DG run, i.e.~Euler without any limiter, the scalability curve
exhibits classic strong scaling behaviour.
Low core counts yield speedups, but the performance stagnates for
bigger counts.
The lower the polynomial order, i.e.~the cheaper the tasks, the earlier we enter
the stagnation regime.
Our adaptivity criterion dynamically refines towards the shock or wave gradient.
This induces a critical path along the refinement fronts which consists of
inter-grid transfer operators and the actual refinement criterion.
If the STP tasks are heavy, we succeed in hiding all this path.
Limiting ADER-DG cures this strong order-dependence as the FV cells yield very
heavy tasks.

\subsection{MPI+X scaling}

%
% What do we do
%
We close this section with MPI+X scaling tests.
To obtain unbiased comparisons, we benchmark against
shared-memory parallelisation only as long as we stay on one compute node.
Shared-memory experiments lack MPI overhead.
Furthermore, we disable all dynamic load balancing, i.e.~we determine a
reasonable domain decomposition pattern prior to the \added{measurements' start}
and stick to this splitting from thereon. 
The splitting uses a uniform cost model,
i.e.~cost per cell, which does not take imbalances into account that arise when
we solve nonlinear PDEs such as the Euler equations with ADER-DG or 
when we apply a localised limiter in the immersed boundary tests.
\added{
 It is a sole geometric decomposition following the Peano space-filling curve
 \cite{Bungartz:06:Parallel,Weinzierl:19:Peano}.
}

\added{
 The data in Table \ref{table:cost-per-dof-per-phase} highlight that 
 a uniform cost model is a particular crude approximation for the  
 limited ADER-DG setup, i.e.~the immersed boundary
 case, since the limiter's FV scheme is by at least one
 order of magnitude more expensive than sole ADER-DG.
 Consequently, we may assume that the load decomposition here is unbalanced.
 Our results however do not suffer qualitatively from this, as we use, per
 experiment, a fixed number of up to 82 MPI ranks, determine a reasonably
 balanced static domain decomposition first, and then increase the number of cores per rank to
 increase the total core count.
 Furthermore, we only benchmark across few time steps, i.e.~the load
 distribution does not shift drastically. 
 The limiter however does stress enclave tasking, as almost all runtime is spent
 on the FV cells.
 They cover only a small subset of the domain.
 Enclave tasking thus has only limited freedom to exploit and hide enclave
 work.
}
%Obviously, this induces imbalances if we use the Euler equations in combination with AMR.
%For Euler, it even yields imbalances for regular grids, as the
%limited region, i.e.~the cells with high cost, move through the domain.
%The approach is slightly imbalanced for the immersed boundary tests as it
%works with a cost model that assumes a uniform cost per cell.
%---it counts the number of ADER-DG and
%FV cells---

\begin{figure}[htb]
  \begin{center}
   \includegraphics[width=0.325\textwidth]{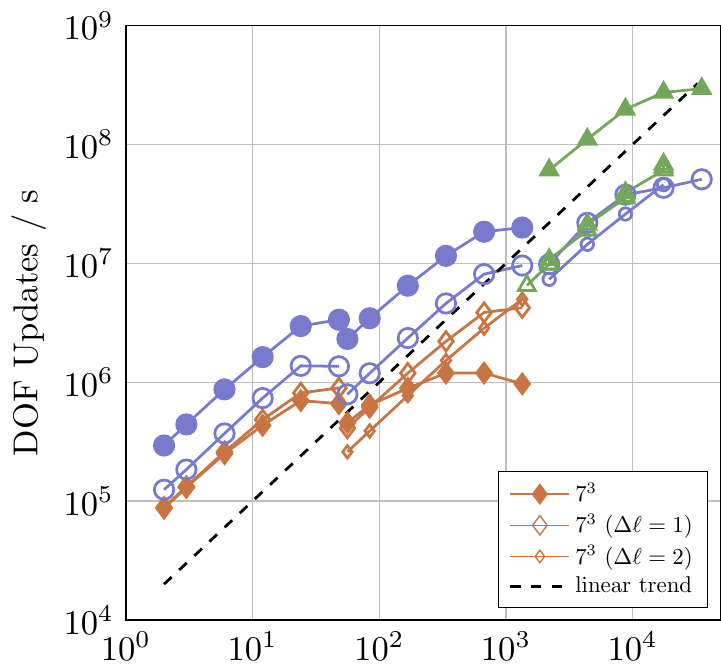}
   \hspace{1.2cm}
   \includegraphics[width=0.325\textwidth]{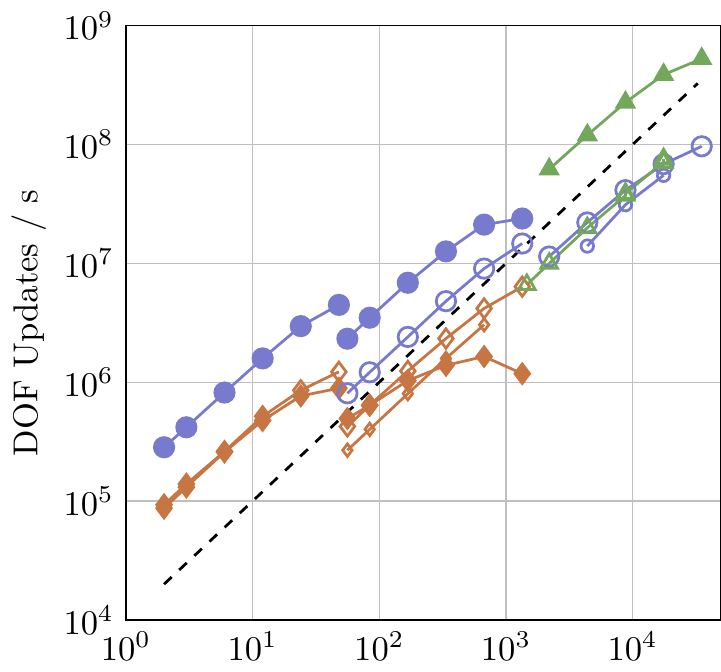}
   \\
   \includegraphics[width=0.325\textwidth]{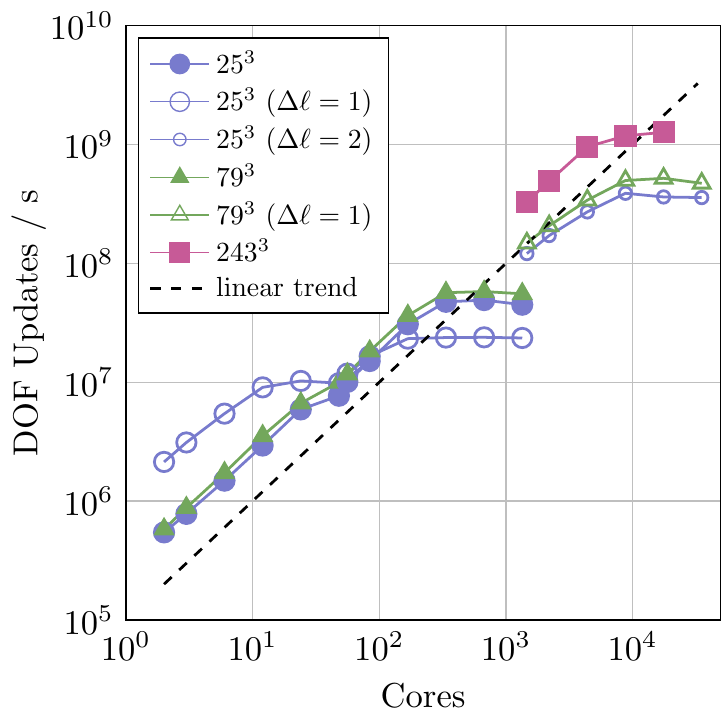}
   \hspace{1.2cm}
   \includegraphics[width=0.325\textwidth]{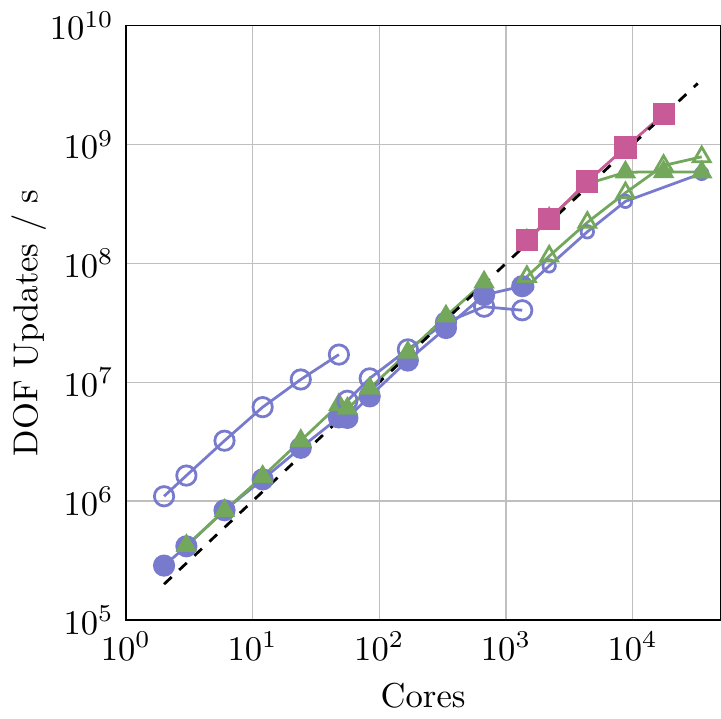}
  \end{center}
  \vspace{-0.4cm}
  \caption{
   MPI+X scaling for the seismic setup (top) and the Euler ADER-DG solver without limiter. We study $p=5$ (left) and $p=7$ (right).
   %See the plots on the left for the legend.
   \added{
    Each individual connected line represents one strong scaling
    experiment.
   }
   \label{figure:results:distributed-memory}
 %\vspace{-0.6cm}
  }
 \end{figure}

%
% What do we see
% 
Both the seismic and the
Euler (Fig.~\ref{figure:results:distributed-memory}) runtimes suffer from
the switch from a shared memory experiment to an MPI+X run.
Each individual problem size yields a strong scaling curve, i.e.~a curve that
starts to stagnate or even deteriorate from a certain core count on, while
we cannot really make a statement alike ``AMR scales less reasonable than its
regular counterpart''.
% The curves exhibit some small stagnation areas or even performance drops for
% certain core counts but otherwise exhibit reasonably robust scaling.
For all solvers, the switch to 
%a higher resolution, i.e.~finer mesh, 
a finer mesh improves the throughput.
Though higher order computations yield more science per degree of freedom,
the rough cost per degree of freedom update is indistinguishable from a
low-order counterpart.

%
% Interpretation
%
The MPI+X executable suffers from overhead, such as MPI
polling cost or global time step synchronisation, compared
to its single-node cousin.
We consequently see a performance drop once we leave the single node.
It is difficult to reconstruct where the other non-smooth effects come from,
but it is reasonable to assume that it stems partially from ill-balancing. 
This comprises not only ill-balancing as discussed above but also certain
core/node counts that do not map perfectly to a given mesh.
The deterioration for too high core counts and the otherwise 
good scaling highlight that enclave tasking automatically
hides communication behind computation.
If the amount of work on a node becomes too small, hiding fails.
The break even point is reached quickly: 
Cells adjacent to the MPI boundary are skeletons and the enclave size thus tends
to shrink drastically relative to the skeleton cardinality once we make the
subdomains smaller.
Besides standard overhead arguments---we obviously also have to invest compute
resources into the mesh management and traversal---this relative growth of
skeleton vs.~enclave size explains why the throughput improves drastically
whenever we increase the mesh resolution.
If we compare a regular mesh to an AMR mesh, 
our AMR results are qualitatively similar to the regular mesh data in most
cases.
This emphasises that we hide both data exchange and AMR inter-resolution
projections successfully behind the enclave work. 
% Along the lines of the relative skeleton vs.~enclave sizes, AMR however yields
% more unfavourable ratios and thus lower throughput; an effect typically
% easily compensated by an overall lower degree of freedom count per accuracy.
Overall, we have successfully transferred the advantageous characteristics of
enclave tasking from the shared-memory into the MPI+X domain.

\section{Conclusion}
\label{section:conclusion}

%
% What have we seen and why. It is a method.
%
Enclave tasking is a powerful technique to equip DG
codes with high concurrency and advantageous communication characteristics.
It removes multicore synchronisation points, overlaps computation and
communication, and it implicitly orchestrates a well-blended mix of compute-intense and memory-intense tasks.
This makes it a powerful tool in particular in the context of dynamic AMR.
Here, it it provides an orthogonal
technology to domain decomposition that increases the parallelism in an MPI+X
environment.

%
% Shortcomings
%
There are conceptional and implementation shortcomings of the present approach.
It firstly performs best in cases where the
adaptivity is localised and the regular subdomains host a sufficient number of
cells.
For extremely high order codes which often host only few cells per node, this
might not hold.
Secondly, our data showcases that it relies on a code where cell tasks
are significantly more expensive than all other tasks.
For some setups, this is only the case for large polynomial orders, for
nonlinear solves within the ADER-DG cells, and for simplistic Riemann solves.
Thirdly, our ``fire-and-forget'' strategy for skeleton and enclave
STP tasks is appropriate for explicit time stepping where the
admissible time step size is well-known or can be estimated.
An extension to local time stepping, implicit time stepping or elliptic problems
where equation systems are solved is beyond scope here.
It certainly requires further work.
Fourthly, our case studies do not invest
into performance engineering or load balancing.
Proper application of these techniques certainly will change all outcomes
quantitatively.

%
% What do we learn for the implementation tools.
%
Enclave tasking's success hinges on the task system.
Our realisation manually adds priorities to Intel's TBB as we found TBB's
native priority scheduling insufficient. 
It also manually polls MPI---to allow the
message exchange to make progress---and it throttles the number of
background tasks.
The processing of enclaves thus never grabs too many cores from the actual main
traversal unless there is an enormous number of STP tasks ready.
We expect the next generation of task runtimes to provide
appropriate support for priorities.
This flavour of our wrapper thus will become obsolete.
All other features are, to the best of our
knowledge, not yet on any task runtime's roadmap.
% Moreover, our experience suggests that future task systems would benefit from
% MPI triggers (guards):
% Tasks can become ready once the MPI subsystem flags
% that there are messages ready to arrive or can poll MPI upon request.
% In theory, such a feature should be supported through hardware.
% It would release us from the responsibility to poll MPI ourselves.
Moreover, our experience suggest that it might be reasonable to equip tasks
with meta flags indicating whether they are bandwidth or compute intense.
A good runtime then can ensure that the bandwidth-intense tasks dribble through
the system, yet, that never too many of these bandwidth-demanding tasks are executed at the
same time.
% With explicit task graphs, such features can be modelled already.
% In the totally dynamic AMR case, memory annotation to obtain the right blend
% of tasks is missing.
% On the long term, they might determine whether our task-free approach is
% superior to classic approaches assembling (parts of) the task graph.
We have made good experience with fusing our cheap Riemann solves with the
traversal.
Memory-expensive inter-grid transfer tasks along resolution boundaries however
do not yet benefit from parallelisation---we mangle them into the traversal,
too---even though we have learned that they tend to align along the critical path.

% %
% % Three follow-up steps.
% %
There are natural follow-up steps and follow-up questions worth further
investigation:
First, enclave tasking is a promising candidate to be used in connection with
accelerators
\cite{Dosopoulos:2011:MPIAndGPU,Ferreira:2017:LoadBalancingAndPatchBased,Godel:2010:Scalability,Heinecke:2014:PetascaleDynamicRupture,Komatitsch:2010:HighOrder,Mu:2013:Accelerating,Klockner:09:DGonGPUs}.
Our terminology is inspired by the work of Sundar and Ghattas \cite{Sundar:15:Enclave}
who use enclaves to ensure that accelerators processing an enclave do not have
to communicate with other accelerators directly.
The enclaves are separated by skeleton cells.
Our background tasks are perfect candidates to be deployed to
accelerators, too.
Their data transfers can be hidden behind computation, and the construction of
the skeleton mesh ensures that no accelerator has to exchange DG jumps with
another enclave.
Second, enclave tasking alters the scaling behaviour of the code base and makes
it depend on the grid topology.
Future work will have to study whether grid refinement criteria should
anticipate this scaling behaviour.
It is reasonable to assume that sophisticated criteria optimise both towards an
as small as possible grid and a scaling grid topology.
\added{
 Their interaction with dynamic load balancing beyond the simple space-filling
 curve cuts employed here \cite{Bungartz:06:Parallel} however is not obvious.
 It is also not clear to which degree generic load balancing strategies can
 succeed or whether good strategies have to incorporate application specifics
 and knowledge.
} 
Finally, enclave tasking has to be studied in the context of
single-sided MPI or distributed shared memory systems where much of the MPI progression pain fades
away.
It is a particularly fascinating idea to study the deployment of enclave
tasks to remote nodes rather than only local cores. 
As enclave tasking adds an additional dimension of concurrency to classic domain
decomposition, this idea adds an orthogonal dimension to classic load balancing.
First studies along these lines are promising \cite{Samfass:19:Offloading}.

\section*{Acknowledgements}
The authors appreciate support received from the European Union’s Horizon 2020
research and innovation programme under grant agreement No 671698 (ExaHyPE).
Our development made use of the facilities of the Hamilton HPC Service of Durham
University. Thanks are due to all members of the ExaHyPE consortium.
All underlying software is open source \cite{Software:ExaHyPE}.

\appendix

\bibliographystyle{plain}
\bibliography{paper}

\end{document}